\documentclass{aa} 
\usepackage{graphicx}
\usepackage[super]{nth}
\usepackage{natbib}
\usepackage{float}
\usepackage{fix-cm}
\usepackage{amsmath}

\let\oriupbracefill\upbracefill
\let\oridownbracefill\downbracefill

\usepackage{txfonts}

\let\upbracefill\oriupbracefill
\let\downbracefill\oridownbracefill

\usepackage{mathtools}
\usepackage{esint}
\usepackage{subfig}

\usepackage{multirow}
\usepackage[percent]{overpic}
\usepackage{xcolor}
\graphicspath{{./figures/}}
\bibpunct{(}{)}{;}{a}{}{,} 
\usepackage{xcolor,pict2e}
\usepackage{float}
\begin{document}

        \title{Interaction of coronal mass ejections and the solar wind. A force analysis}
        
        \author{D.-C. Talpeanu \inst{1,2}
                \and
                S. Poedts \inst{1,3}
                \and
                E. D'Huys\inst{2}
                \and
                M. Mierla \inst{2,4}
                \and
                I.G. Richardson \inst{5,6}
        }
        
        \institute{Centre for mathematical Plasma Astrophysics (CmPA), Department of Mathematics, KU Leuven, Celestijnenlaan 200B, 3001 Leuven, Belgium\\
                \email{dana.talpeanu@observatory.be}
                \and
                SIDC - Royal Observatory of Belgium (ROB), Av. Circulaire 3, 1180 Brussels, Belgium               
                \and
                Institute of Physics, University of Maria Curie-Sk{\l}odowska, Pl.\ M.\ Curie-Sk{\l}odowska 5, 20-031 Lublin, Poland
                \and 
                Institute of Geodynamics of the Romanian Academy, Jean-Louis Calderon 19-21, 020032 Bucharest, Romania
                \and
                Department of Astronomy, University of Maryland, College Park, MD 20742, USA 
                \and
                Heliophysics Division, NASA Goddard Space Flight Center, Greenbelt, MD 20771, USA      
        }
        
        \date{Received : to be filled in; accepted: to be filled in}
        
        % \abstract{}{}{}{}{} 
        % 5 {} token are mandatory
        
        \abstract
        {}
        {Our goal is to thoroughly analyse the dynamics of single and multiple solar eruptions, as well as a stealth ejecta. The data were obtained through self-consistent numerical simulations performed in a previous study. We also assess the effect of a different background solar wind on the propagation of these ejecta to Earth.}
        {We calculated all the components of the forces contributing to the evolution of the numerically modelled consecutive coronal mass ejections (CMEs) obtained with the 2.5D magnetohydrodynamics (MHD) module of the code MPI-AMRVAC. We analysed the thermal and magnetic pressure gradients and the magnetic tension dictating the formation of several flux ropes in different locations in the aftermath of the eruptions. These three components were tracked in the equatorial plane during the propagation of the CMEs to Earth. Their interaction with other CMEs and with the background solar wind was also studied.
        }
        {We explain the formation of the stealth ejecta and the plasma blobs (or plasmoids) occurring in the aftermath of solar eruptions. We also address the faster eruption of a CME in one case with a different background wind, even when the same triggering boundary motions were applied, and attribute this to the slightly different magnetic configuration and the large neighbouring arcade. The thermal pressure gradient revealed a shock in front of these slow eruptions, formed during their propagation to 1~AU. The double-peaked magnetic pressure gradient indicates that the triggering method affects the structure of the CMEs and that a part of the adjacent streamer is ejected along with the CME.}
        {}
        
        \keywords{magnetohydrodynamics (MHD) --
                methods: numerical --
                Sun: coronal mass ejections (CMEs)
        }
        
        \maketitle
        
        %%-------------------------------------------------------------------
        
\section{Introduction} \label{sec:intro}
        
Coronal mass ejections (CMEs) belong to the most frequently encountered and energetic large-scale solar phenomena through which up to 10$^{16}\;$g \citep{schwenn_review, chen_review_cmes} of magnetized plasma are ejected into interplanetary space. The frequency of these eruptions based on spacecraft coronagraph observations varies greatly throughout the solar cycle, from $\approx$ 0.5 CMEs per day during solar minimum to 5-6 CMEs per day during solar maximum \citep{cactus, lamy_2019}. CME interactions have been observed near the Sun and in interplanetary space. Using remote-sensing observations and in-situ data, such interactions between two successively launched CMEs have been tracked and analysed from the Sun to Earth by several authors, such as \citet{manuela_cme_cme_interaction2012} and \citet{manuela_cme_cme_interaction2014}, and all the way to Saturn by \citet{erika_cme_cme_interaction}. The launch of the twin Solar Terrestrial Relations Observatory (STEREO) spacecraft in 2006 \citep{stereo_mission} has prompted  studies \citep[e.g.][]{liu_2014, webb_2013} of successive CMEs using the multi-viewpoint capabilities provided by instruments on these spacecraft, in combination with in-situ data. \citet{liu_2014} also analysed two CMEs interacting near 1~AU that resulted in a two-step geomagnetic storm driven by their complex ejecta.\\
The mechanisms by which ICMEs interact have been investigated using numerical simulations by several authors, such as \citet{torok_sympathetic_eruptions}, who modelled two sympathetic eruptions based on an observed event; \citet{lugaz_successive_cmes}, who simulated the interaction of two CMEs from the Sun to Earth; and \citet{bemporad}, who performed magnetohydrodynamics (MHD) simulations to physically explain the origin of multiple CMEs from an asymmetric coronal field configuration. \\    
The morphology and dynamics of CMEs also vary with the solar activity cycle. CMEs are wider and faster during solar maximum, and they are usually associated with other features such as flares, coronal dimmings, filament eruptions, extreme-ultraviolet (EUV) waves, or post-flare loops \citep{hudson_cme_sources}. All these features are good indicators of the source region of the eruptions. During solar minimum, the percentage of CMEs associated with such signatures is much lower \citep{ma_stealth}, leading to extreme cases, so-called stealth CMEs \citep{robbrecht_stealth, nariaki_stealth, nariaki_stealth_review, lynch_stealth}, that are observed in coronagraph images, but are difficult to trace back to their origin on the Sun. \citet{robbrecht_stealth} concluded that one reason why these ejections do not present clear source signatures is their large lift-off heights, usually above 1.4 solar radii (R$_{\sun}$). Even though their sources are difficult to locate, stealth CMEs can be unexpectedly geoeffective \citep{kilpua_stealth, liu_2016, nariaki_stealth_review}. One reason is that the magnetic fields of slow and stealth CMEs may be enhanced by interactions with the background solar wind through which they are propagating, resulting in surprisingly strong geomagnetic storms \citep{he_2018_stealth, chen_2019}.\\
\citet{talpeanu_paper1} (referred to here as paper~I) performed self-consistent 2.5D MHD numerical simulations of consecutive CMEs obtained by applying time-dependent shearing motions at the inner boundary of the computational domain, along the southernmost polarity inversion line of a triple arcade magnetic structure. One of the eruptions was a blob-like stealth ejecta, occurring in the trailing current sheet created by the passage of a first CME. In a follow-up study, \citet{talpeanu_paper2} (paper~II) investigated the effect of a faster and denser background solar wind on the initial magnetic configuration and the resulting eruptions obtained through the same triggering mechanism. Firstly, they reported a split of the initial overlying streamer into a northern arcade and a southern pseudostreamer. Secondly, when applying the same shearing speed amplitude as in paper~I, the flux ropes formed earlier and the associated CME had a higher speed in the faster solar wind case. Lastly, the stealth ejecta no longer formed, nor did the plasma blobs (or plasmoids) following the main eruptions in the trailing current sheet.\\                     
It is known that the upward magnetic and plasma pressure gradients, balancing the downward magnetic tension and gravity, keep magnetic structures and filaments in equilibrium in the solar atmosphere \citep{chen_1996, forbes_energies, schmieder2013}. The exact interplay and imbalance between all these forces that eventually drive CMEs has not been extensively documented in the literature. \citet{cargill_2002} and \citet{cargill_2004} investigated the forces acting on numerically simulated CMEs during propagation, as well as the coupling between the CMEs and the background solar wind expressed through the aerodynamic drag \citep[e.g.][]{chen_garren_1993,cargill_1995}. The few more recent studies of these physical processes include \citet{shen_forces}, who analysed the forces causing acceleration and deceleration of CMEs, and \citet{skralan_forces}, who studied the deformation and erosion of CMEs during their evolution. Furthermore, \citet{kay_2021} presented the first results of their numerical code that simulates the propagation, expansion, and deformation of a CME in the interplanetary medium, and quantified the contributions of each of the imposed forces to these processes. Closer to the Sun, \citet{lynch_2009} explain the rotation of a CME during eruption by analysing the forces involved. The review of \citet{manchester_cmes_interaction} summarises the ways in which CMEs and their interplanetary counterparts evolve during their journey through the solar wind.\\ 
In order to thoroughly explain the differences in dynamics observed in papers I and II, we calculate all the contributing forces that govern the resulting eruptions. We also investigate the CME-CME and CME-solar wind interactions during their propagation away from the Sun.    
The numerical set-up and the methods used are described in Section~2. The detailed force analysis during CME onset and evolution is provided in Section~3. Finally, Section~4 contains the summary and our conclusions.
        
        %%-------------------------------------------------------------------   
        
\section{Numerical MHD code, methods, and simulations} \label{sec:simulations}
        %\subsection{Simulation setup} \label{subsec:sim_setup}
        
\subsection{Numerical setup} \label{subsec:numerical_setup}
        
The goal of this follow-up study is to analyse and understand the forces dominating the CMEs simulated and described in paper~II from their eruption phase to their propagation in interplanetary space. Because the numerical setups are identical to those used in paper~II, we limit ourselves here to a few remarks about the code. If the results are to be reproduced, the mathematical description of the background solar wind, initial conditions, and CME triggering method can be found in paper~II.\\         
The numerical MHD simulations we analysed were performed using the Message Passing Interface - Adaptive Mesh Refinement - Versatile Advection Code \citep[MPI-AMRVAC;][]{keppens_amrvac, porth_amrvac, xia_amrvac}, with a 2.5D spherical axisymmetric geometry. The computational mesh extends from the low corona until $1.5\;$AU, and from the north solar pole to the south solar pole, that is, (\emph{r}, $\theta$) $\in$ [1, 322] R$_{\sun}$ $\times$ [0\degr, 180\degr], with $r$ being the radial distance from the centre of the Sun, and $\theta$ is the heliographic colatitude. The initial resolution of the 2D logarithmically stretched grid contains 516 $\times$ 240 cells in the $r$ and $\theta$ directions, and this number increases dynamically up to two times via an AMR routine that traces the current-carrying structures that potentially appear at magnetic reconnection sites. This method refines the cells in both directions through a parameter $c$ that can have values between 0 and 1, depending on the magnetic field and its non-potentiality. This parameter has previously been used by \citet{karpen} and \citet{skralan}, and also in the simulations performed in papers I and II. The AMR criteria are the following: if $c > 0.02$ in a block of cells, then the grid is refined in that particular block; if $c < 0.01$, then the grid is coarsened, since there are no strong current-carrying structures in that region; if $c$ has a magnitude between these two values, then the grid remains at the resolution imposed by the AMR routine in the previous time step. These conditions ensure that the high-resolution mesh tracks the CME and the current sheets. An additional fixed maximum refinement is imposed close to the Sun, in an area that encompasses the coronal magnetic structures, with the purpose of ruling out diffusivity changes and artificial dynamics during the initiation phase of CMEs. This constant region is dependent solely on the background solar wind used for simulations, and not on the parameter $c$. The scale of the cells is kept constant through a logarithmic stretching method, and the ratio of the widths or heights of the furthest cell to the closest cell belonging to the same grid level of refinement is $\approx$321.\\    
The grid cells split the computational domain into well-determined intervals, but the MHD equations are defined in a continuous manner. They therefore need to be discretized spatially as well as temporally. The methods used to accomplish these divisions are the total variation diminishing Lax-Friedrichs (TVDLF) scheme and a two-step predictor corrector, respectively. A very stable and diffuse configuration was achieved by using the slope limiter minmod in combination with a Courant-Friedrichs-Lewy (CFL) number of 0.3. The method used to keep the magnetic field solution divergence free is called the generalised Lagrange multiplier \citep[GLM; ][]{glm}, which smooths and transports to the outer boundary of the computational domain the unphysical monopoles that might be created.\\  
The CMEs are erupted into a bimodal background solar wind symmetric in the $\phi$ direction due to the 2.5D (axisymmetric) geometry, which was obtained by introducing additional source terms to the momentum and energy equation that account for gravity and heating mechanisms. The bimodality consists of a latitudinal dependence of the solar wind speed, with faster solar wind close to the poles, consistent with the data obtained by the Solar Wind Observations Over the Poles of the Sun \citep[SWOOPS;][]{swoops_ulysses} instrument \citep{mccomas_solar_wind} on board the Ulysses spacecraft \citep{ulysses_solar_wind}, as well as with interplanetary scintillation \citep[IPS,][]{interplanetary_scintillation, kojima_ips, rickett_ips} observations. This model of the solar wind has previously been used by \citet{jacobs_2005}, \citet{chane_2006}, \citet{chane_2008}, \citet{skralan_2019}, and in papers I and II, where the empirical volumetric heating function that defines the separation between slow and fast wind is explained as well \citep{groth, manchester}. The simulations were performed at two different background solar wind speeds and densities, which we refer to as slow wind and faster wind. These terms denote two separate configurations, rather than composite latitudinal parts of the same solar wind. The simulated maximum and minimum speeds of the two configurations at 1 AU are the following: 735 km s$^{-1}$ and 330.6 km s$^{-1}$ for the slow-wind case (they resemble the values recorded at Earth for an observed event studied in papers I and II), and 786.3 km s$^{-1}$ and 375.7 km s$^{-1}$ for the faster-wind configuration. The maximum speeds are found at the north pole (90$\degr$ latitude) and the minima in the equatorial current sheet, which is shifted northward due to the initial asymmetric magnetic configuration described below.\\          
In these simulations, the mass density was fixed at the inner boundary to $1.66 \times 10^{-16}\,\mathrm{g \, cm^{-3}}$ for both winds, while the temperature was set to $1.32 \times 10^{6}\,\mathrm{K}$ for the slow wind and to $1.5 \times 10^{6}\,\mathrm{K}$ for the faster wind. The radial component of the momentum was extrapolated in the ghost cells, and at the inner boundary, the latitudinal component of the speed ($v_\theta$) was set to 0. The 2.5D geometry allows calculating a $\phi$ component of vectors. Therefore, the differential rotation of the Sun is reproduced by introducing an azimuthal component of the speed ($v_\phi$). The dipole magnetic field of the Sun was created by fixing $r^2B_r$ at the inner boundary and by extrapolating $r^5B_\theta$ and $B_\phi$ from the first inner cell. The variables $r^2\rho$, $r^2\rho v_r$, $\rho v_\theta$, $r v_\phi$, $r^2B_r$, $B_\theta$, $r B_\phi$, and $T$ are also continuous at the outer supersonic boundary, which lies at $1.5\;$AU.\par

\begin{figure}[h!]
        \centering
        \begin{overpic}[width=1\columnwidth]{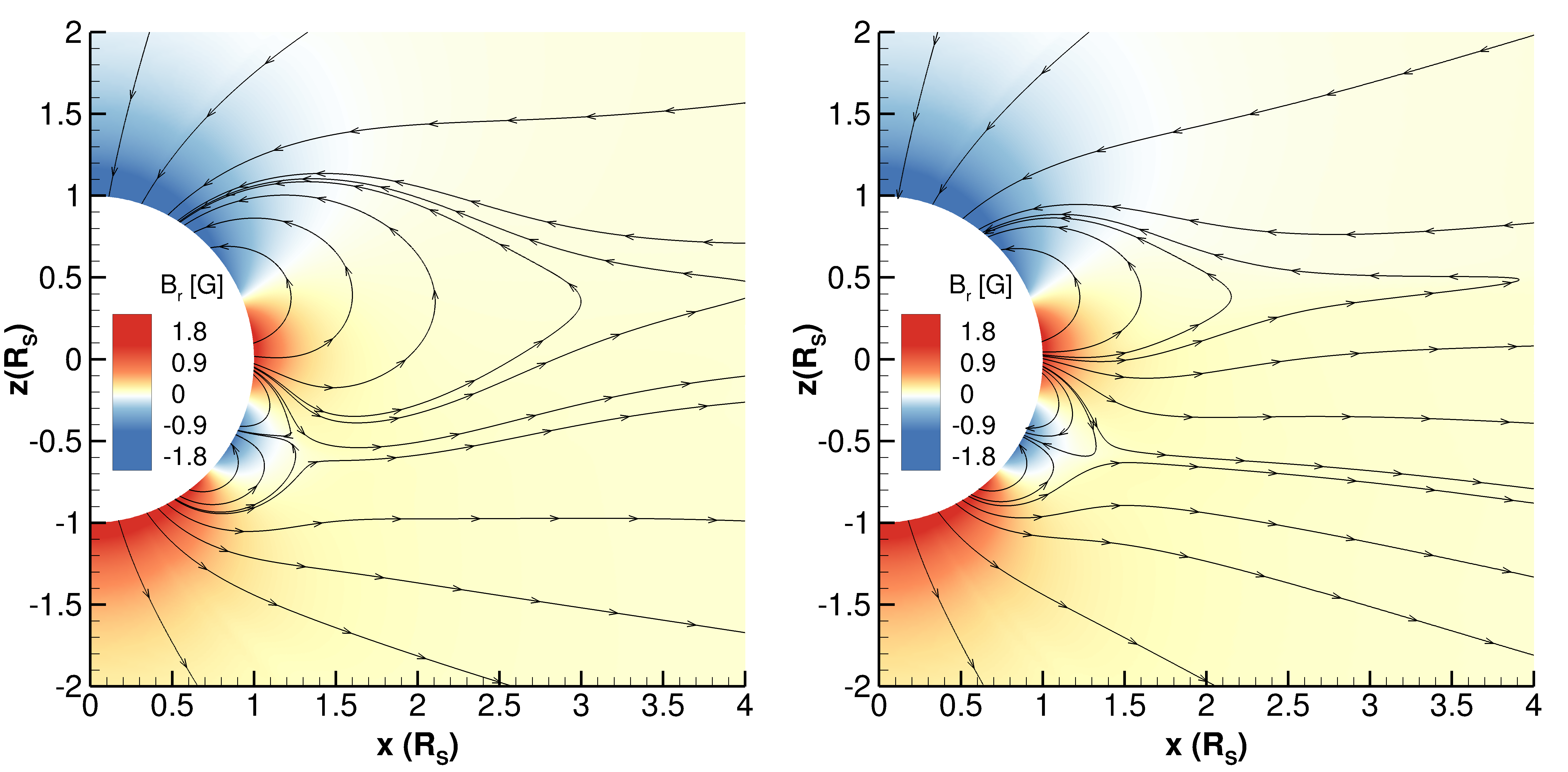}
            \put(17.5,15.5){\color{black} \vector(-1,1){4}}
            \put(67.7,15.5){\color{black} \vector(-1,1){4}}
        \end{overpic}    
        \caption{Initial magnetic field configuration - $B_r$ (colour scale) and selected magnetic field lines in the meridional plane for the slow solar wind configuration (left side) and faster solar wind configuration (right side). The large black arrows indicate the position of the southernmost polarity inversion line in both configurations. Figure reproduced with permission from \citet{talpeanu_paper2}.}
        \label{fig:arcades}     
\end{figure}

In addition to the dipole field, the coronal magnetic structure for the slow-wind case depicts an asymmetric triple arcade system embedded in a helmet streamer, as previously simulated and used by \citet{bemporad} and \citet{zuccarello} to analyse the dynamics and deflection of a multiple eruption event, and by \citet{karpen} to study a breakout event. The arcades were shifted southward by 11.5$\degr$ and were added between latitudes $[-48.7\degr, 25.8\degr]$. The two small arcades in the southern hemisphere constitute a local pseudostreamer, but due to its proximity to the large northern arcade, the pseudostreamer tends to follow the boundary of the northern arcade and helmet streamer. Thus, the pseudostreamer may be considered as part of a combined helmet streamer. When the momentum was increased in order to obtain the faster solar wind configuration, the helmet streamer broke up into the northern arcade and the southern pseudostreamer, without modifying the magnetic field boundary conditions. The pseudostreamer was then considered a separate entity since it is directed almost radially away from the Sun and is less influenced by the northern structure. These two simulated configurations are shown in Fig.~\ref{fig:arcades}. They provide a radial magnetic field strength of 1.8 G (or ${1.8 \times 10^5}$ nT) at the poles and a maximum arcade strength of 1.57 G (or ${1.57 \times 10^5}$ nT). These values are taken from the first cell of the domain.\par 
When the background solar winds reached steady state, CMEs were obtained by applying time-dependent shearing motions along the southernmost polarity inversion line for 16 h in each simulation. The shearing amplitude was kept physical by never exceeding $10\%$ of the local Alfv\'{e}n speed, and the temporal profile presents a slow rise and decrease, with a maximum speed at half the applied time period (at 8 h).    

%%-------------------------------------------------------------------

\subsection{Simulated eruptions} \label{subsec:simulations} 

We analyse here five simulations performed in paper~II (see Fig.~6 in paper II), which are briefly described in this subsection. In all cases, the first erupting flux rope is driven by the boundary shearing motions that increase the magnetic pressure inside the southernmost arcade, causing it to rise and expand. This affects the initial balance between the magnetic tension and the magnetic pressure gradient and compresses the magnetic field locally. The numerical resistivity induces reconnection between the sides of the arcade, creating a flux rope that begins to erupt. Under the influence of the southern polar magnetic pressure, the erupting flux rope is deflected towards the equator and propagates almost radially inside the equatorial current sheet beyond several solar radii; this height depends on the respective scenario. The differences in amplitude of the applied shearing speed result in the occurrence or absence of secondary CMEs. Each simulation we investigated presents the following dynamics (the speeds are taken at the first cell of the domain):            

\begin{enumerate}
        \item Single eruption - slow wind ($|v_\phi^{max}|=21.95 \mathrm{\ km\ s^{-1}}$): The low amplitude of the shearing motions creates only one erupting flux rope 12 h after the start of the shearing. It forms, erupts, and deflects as described above. Several plasma blobs (or plasmoids) arise in the current sheet formed in the wake of the CME. Most magnetically reconnect with their precursor, as they are ejected into a depleted solar wind environment created by the passage of the first eruption.  
        
        \item Eruption + stealth - slow wind ($|v_\phi^{max}|=37 \mathrm{\ km\ s^{-1}}$): After 8 h of shearing, the first erupting flux rope forms due to the boundary condition imposed on the speed. In its trailing current sheet, another flux rope is created from the coronal magnetic field reconfiguration and follows a similar deflected path as its precursor. This second eruption was considered a stealth ejecta due to the large height of the reconnection site (1.4 R$_{\sun}$) and the slow speed of the event, processes which combined would not be associated with strong low-coronal signatures. Before the stealth ejecta forms, another flux rope is created by the applied shearing motions, but the energy input is not sufficient to cause it to detach, so that it falls back onto the Sun. During the propagation to Earth, the stealth ejecta magnetically reconnects with the first CME, and several plasma blobs also arise in the trailing current sheet.  

        \item Double eruption - slow wind ($|v_\phi^{max}|=37.43 \mathrm{\ km\ s^{-1}}$): Since the amplitude of the shearing speed is nearly identical to that applied in the eruption + stealth case, the first CME is basically unaffected by this small change and presents the same dynamics. The second CME, on the other hand, is now also triggered by the shearing motions at the inner boundary and replaces the stealth ejecta. Even though it is wider and has stronger white-light properties than the stealth ejecta, the second CME still reconnects with the first during their propagation to 1AU. In this case, several plasma blobs also occur in the trailing current sheet.   
        
        \item Single eruption - faster wind ($|v_\phi^{max}|=22.33 \mathrm{\ km\ s^{-1}}$): Despite the similar shearing amplitude, a single erupting flux rope forms earlier than in the slow wind - single eruption case, only 10 h after the start of shearing. It also erupts faster, but it still experiences the northward deflection due to the southern polar magnetic pressure. The CME arrives at Earth as one pancaked flux rope, followed only by the trailing streamer current sheet forming in its wake, since no plasma blobs are created in the simulation. 
        
        \item Stealth speed - faster wind ($|v_\phi^{max}|=36.77 \mathrm{\ km\ s^{-1}}$): The amplitude of the shearing speed applied at the boundary is the same as that which resulted in an eruption + stealth in the slow-wind case, hence the name of this scenario. The first and second flux ropes form 7 h and 12 h, respectively, after the start of shearing, and they are both deflected northward into the equatorial current sheet. The second flux rope is created from the southernmost arcade, where the shear is applied. Hence, it is a result of the triggering motions, as opposed to the stealth ejecta, which arises from magnetic reconnection of the trailing current sheet. After only 19 h from the start of shearing, the two CMEs merge and propagate to Earth as one magnetic entity. No plasma blobs occur in this case either, which leads us to conclude that their formation is not dependent on the triggering conditions, but rather on the initial magnetic field configuration.        
        
\end{enumerate}

%%-------------------------------------------------------------------

\section{Force analysis} \label{sec:force_analysis}

In order to explain the evolution and interactions of the simulated eruptions, as well as the formation of different flux ropes in certain situations, we calculated and analysed the individual forces contributing to their dynamics. The mathematical description of the force densities in a unit volume is given by the MHD momentum equation,  

\begin{equation}
\rho \frac{d \mathbf{v}}{dt}=- \nabla P_P +\underbrace{\mathbf{j} \times \mathbf{B}}_{\let\scriptstyle\textstyle \substack{\mathclap{\text{Lorentz force}}}} + \rho \boldsymbol{g} = - \nabla P_P %
\overbrace{\underbrace{- \nabla \frac{|B|^2}{2 \mu_{0}}}_{\let\scriptstyle\textstyle \substack{\text{$-\nabla P_B$}}} +%
         \underbrace{\frac{(\mathbf{B} \cdot \nabla)\mathbf{B}}{\mu_{0}}}_{\let\scriptstyle\textstyle \substack{\text{$\mathbf{T}_B$}}}}^{\let\scriptstyle\textstyle \substack{\text{$\mathbf{j} \times \mathbf{B}$}}}%
        +\rho\mathbf{g}. 
\end{equation}

Each component was calculated throughout the computational domain with their corresponding sign, since that represents their contribution to the total force. The first term on the right-hand side of the equation is the plasma pressure gradient, $- \nabla P_P$. The Lorentz force $\left( \mathbf{j} \times \mathbf{B} \right)$ (in MHD the electric force is ignored) was decomposed into its constituents parts, the magnetic pressure gradient $\left(-\nabla \frac{|B|^2}{2 \mu_{0}} \right)$, which we denote as $-\nabla P_B$, and the magnetic tension $\left( \frac{(\mathbf{B} \cdot \nabla)\mathbf{B}}{\mu_{0}} \right)$, which we write as $\mathbf{T}_B$. The last term $\left( \rho\mathbf{g} \right)$ represents the gravity of the Sun and was added only to the radial component of the force.\\   
We discuss and explain several interesting aspects of the simulations, using the calculated forces. The topics we examined are the following:

%\vspace{-10px} 
\begin{enumerate}
        \item The faster eruption of CMEs in the faster solar wind cases, when the same shearing speed is applied as in the slow wind cases;
        \item The formation of the stealth ejecta;
        \item The occurrence of plasma blobs (or plasmoids) in the aftermath of eruptions, only in the slow wind cases;
        \item The propagation of CMEs and their interaction with the background solar wind and with other CMEs.\\
\end{enumerate}

%%-------------------------------------------------------------------

\subsection{Different CME eruption times for the two background solar winds} \label{subsec:faster cme in faster wind}

\begin{figure}[b]
	\centering
	\includegraphics[width=0.8\columnwidth]{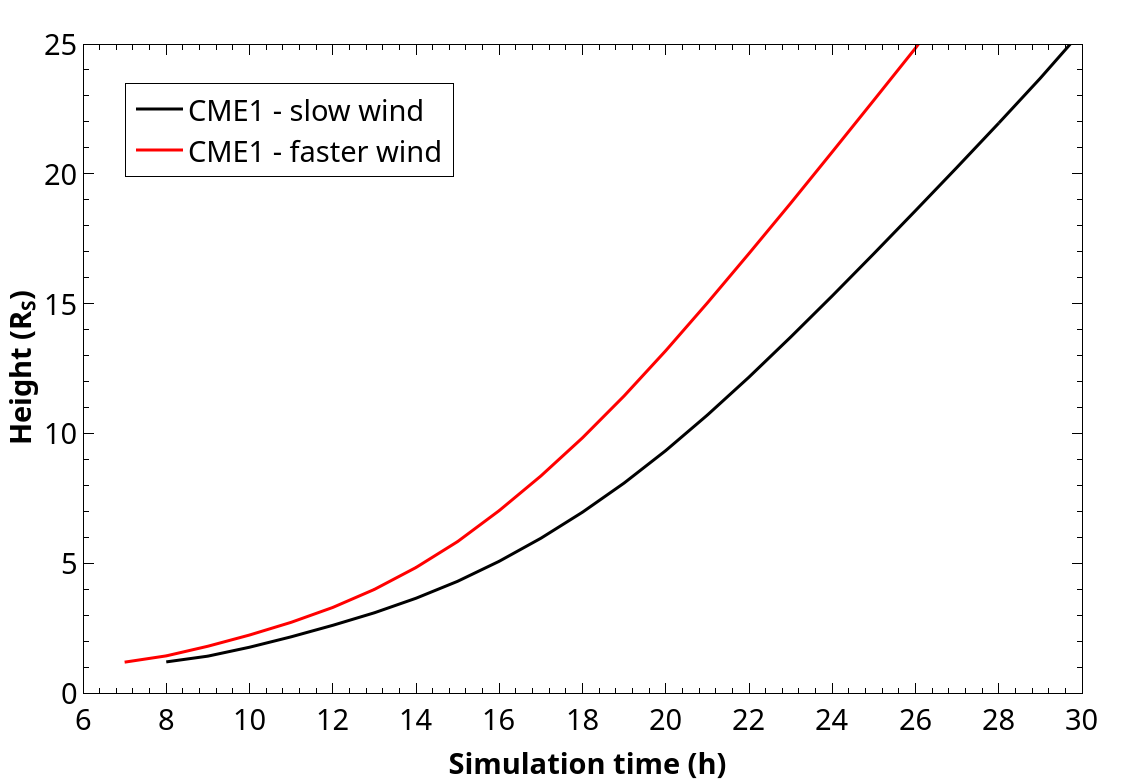}
	\caption{Height-time plot of the centre of the flux rope associated with the first CME in the cases of slow wind - eruption + stealth (black line), and faster wind - stealth speed (red line). The height is calculated from the centre of the Sun, and the simulation time is from the start of the shearing motions.}
	\label{fig:h-t_CME1_both_winds} 
\end{figure}

The first aspect we address is the quicker formation and eruption of CMEs in the faster solar wind cases, when the same shearing is applied as in the slow solar wind cases. As mentioned in Section \ref{sec:simulations}, two shearing speeds of similar amplitudes were applied to the southernmost arcade of both background magnetic field configurations (and solar winds), and yet they resulted in different CME dynamics. The flux rope in the faster wind - single eruption case forms two hours earlier than in the slow-wind case, and the first flux rope in the stealth speed case is created an hour earlier than its equivalent eruption in the slow-wind case (eruption + stealth). This second comparison is shown in Fig. \ref{fig:h-t_CME1_both_winds}, which shows the height of the centre of the flux rope versus time; the steeper slope of the red line indicates a higher eruption speed of CME1 in the faster-wind case. It is also interesting to note that the temporal difference between the equivalent scenarios (between the two background winds) increases as the amplitude of the shearing speed decreases. The simulated shearing motions are intended to represent the accumulated footpoint displacement resulting from weeks of real differential rotation \citep{lynch_stealth}, in addition to local photospheric movements strongly correlated to magnetic flux emergence \citep{wang_shear_motions_flux_emergence}. Our simulations indicate that the weaker these combined motions, the stronger the effect of the coronal magnetic environment on the dynamics of the flux ropes. The lower limit of this assumption would be that at some point, the shearing motions will not produce any eruption in the slow wind case, whereas a CME will still be created in the faster-wind configuration.\par

\begin{figure}
        \centering
        \includegraphics[width=0.5\textwidth]{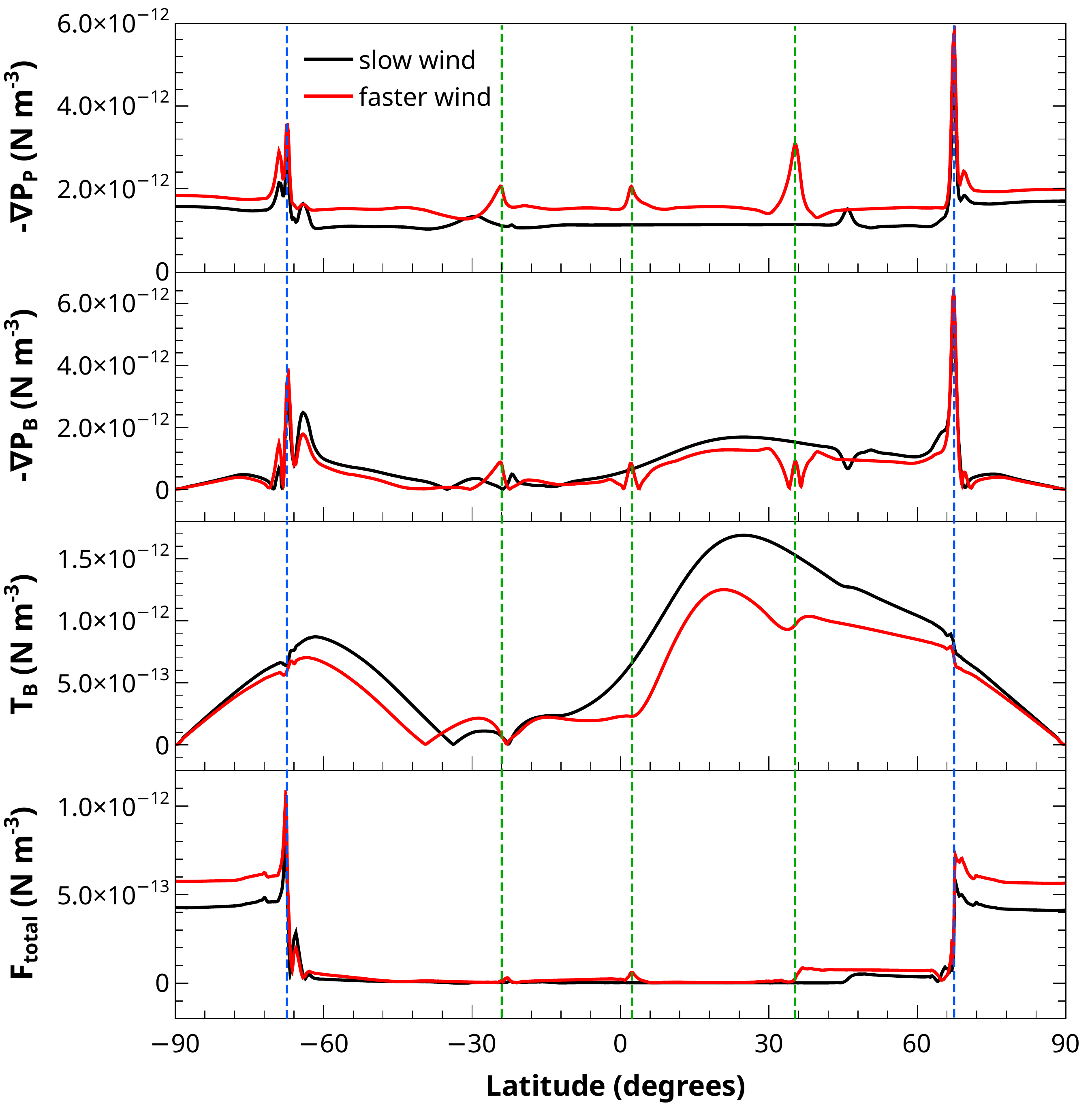}
        \caption{Forces extracted from the 1D cut at 1.5 R$_{\sun}$ (blue semicircles in Fig. \ref{fig:peaks_identification}) in the cases of slow (black line) and faster background solar wind (red line). The force densities (top to bottom) are the following: plasma pressure gradient ($- \nabla P_P$), magnetic pressure gradient ($-\nabla P_B$), magnetic tension ($\mathbf{T}_B$), and total force ($F_{total}$). The dashed blue lines indicate the interface between slow and fast wind, here referred to as latitudinal components of the same background wind simulation. The dashed green lines represent peaks in $- \nabla P_P$ in the faster-wind case (here referred to as the separate simulation case), which is identified by the white circles in Fig. \ref{fig:peaks_identification}.}
        \label{fig:forces_1.5Rs_winds}  
\end{figure}

\begin{figure}
	\centering
	\includegraphics[width=0.48\textwidth]{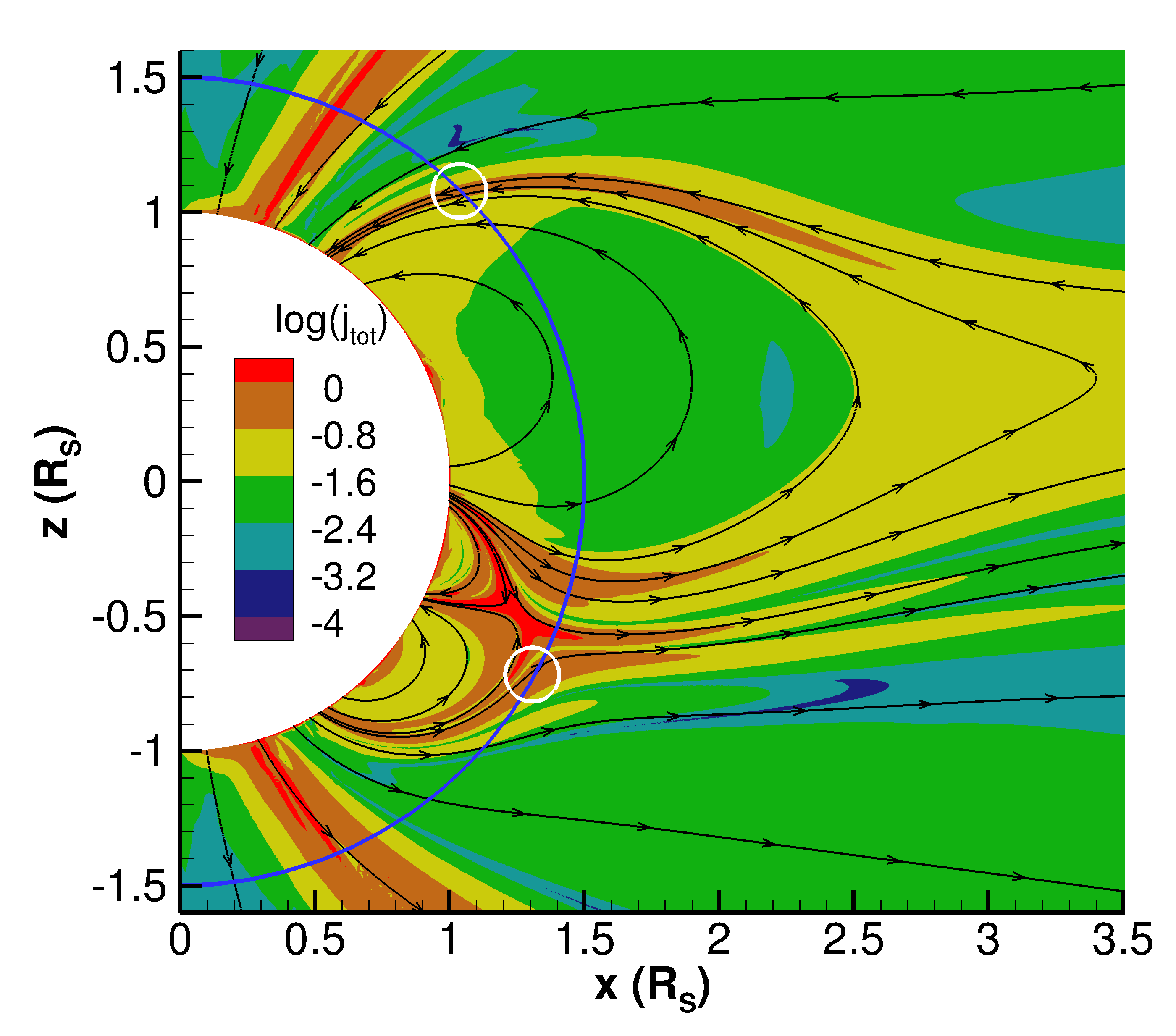}\\
	\vspace{-5px}
	\includegraphics[width=0.48\textwidth]{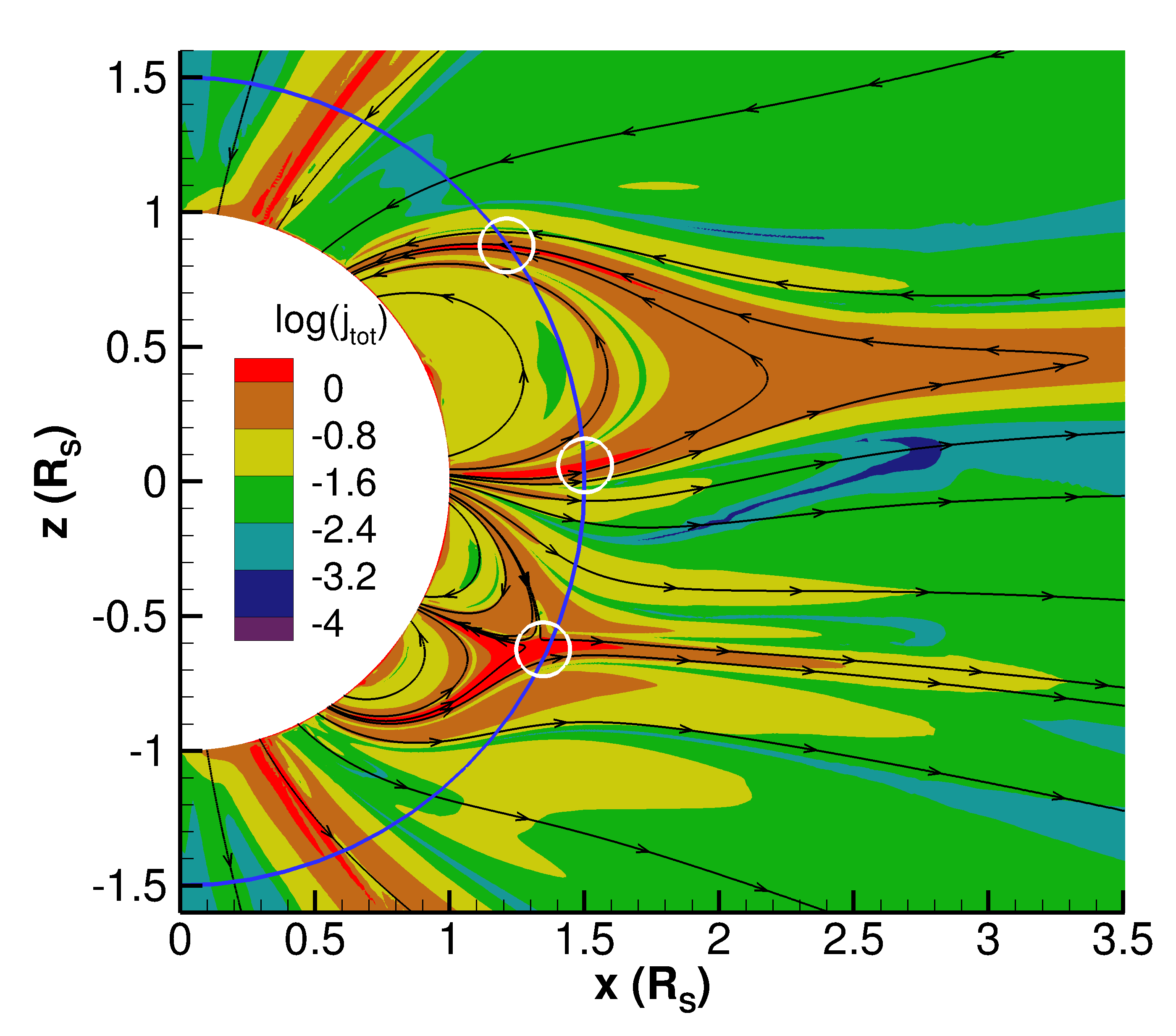}
	\vspace{-10px}
	\caption{Simulation snapshots of the electrical current density (colour scale) in the relaxed slow (top panel) and faster (bottom panel) background solar winds. The black lines represent selected magnetic field lines. The blue semicircles at 1.5 R$_{\sun}$ are the cuts from where the forces in Fig. \ref{fig:forces_1.5Rs_winds} were extracted. The white circles approximately indicate the locations of the $- \nabla P_P$ peaks in Fig. \ref{fig:forces_1.5Rs_winds}.}
	\label{fig:peaks_identification}        
\end{figure}

In order to analyse the results of our simulations, we calculated the forces in the relaxed background wind because since we imposed approximately the same shearing speed, the triggering method cannot be the cause of the different eruption scenarios. We computed the forces in the entire domain for both solar winds and then extracted them at 1.5 R$_{\sun}$ at all latitudes (see Fig. \ref{fig:forces_1.5Rs_winds}). This distance was chosen because it is just above the southernmost arcade and null point. We may therefore obtain the influence of the overlying coronal environment on the eruptions. The blue semicircle in Fig. \ref{fig:peaks_identification} represents the location of the force values plotted in Fig. \ref{fig:forces_1.5Rs_winds}.\par   

\begin{figure*}
	\centering
	\includegraphics[width=0.9\textwidth]{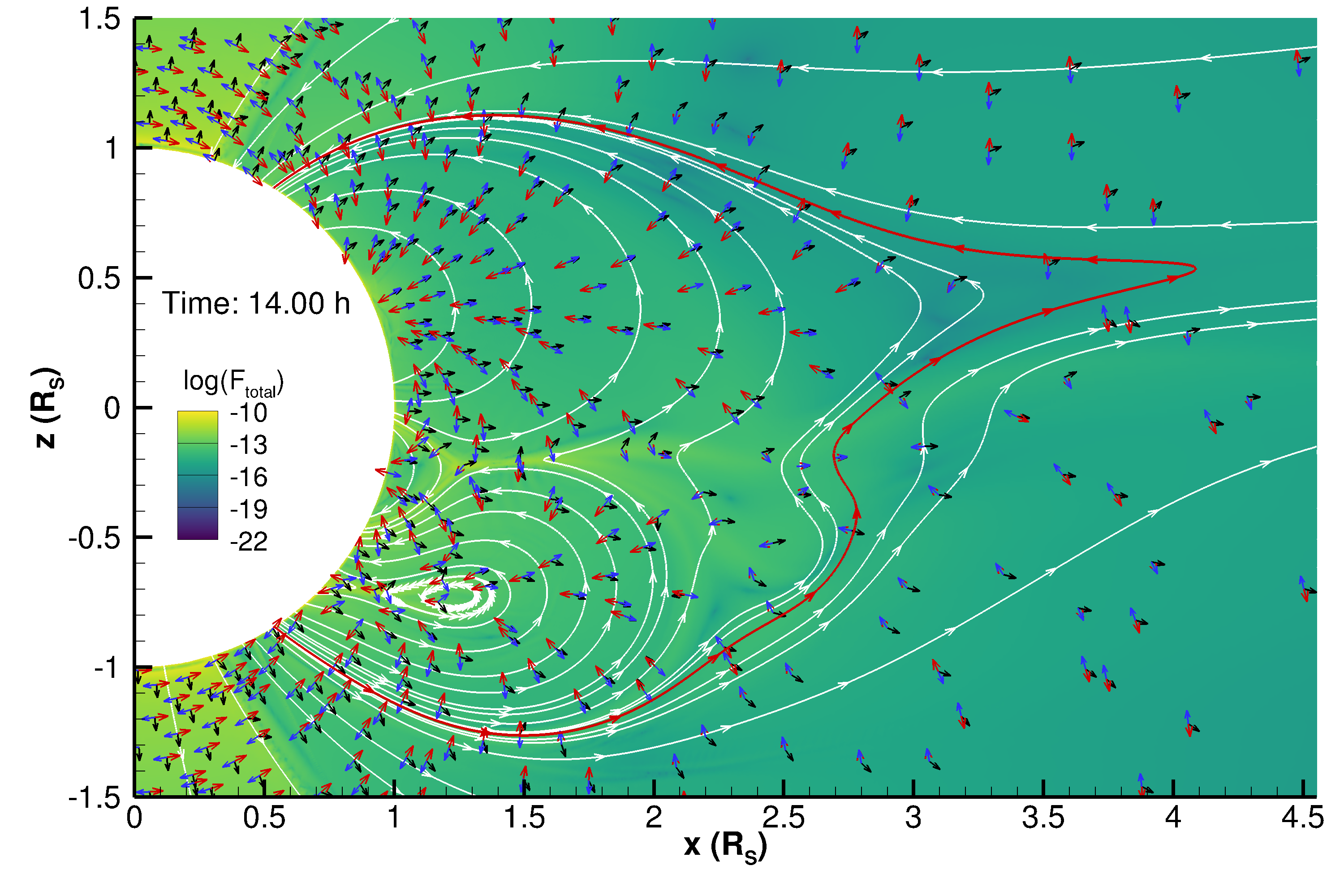}
	\includegraphics[width=0.9\textwidth]{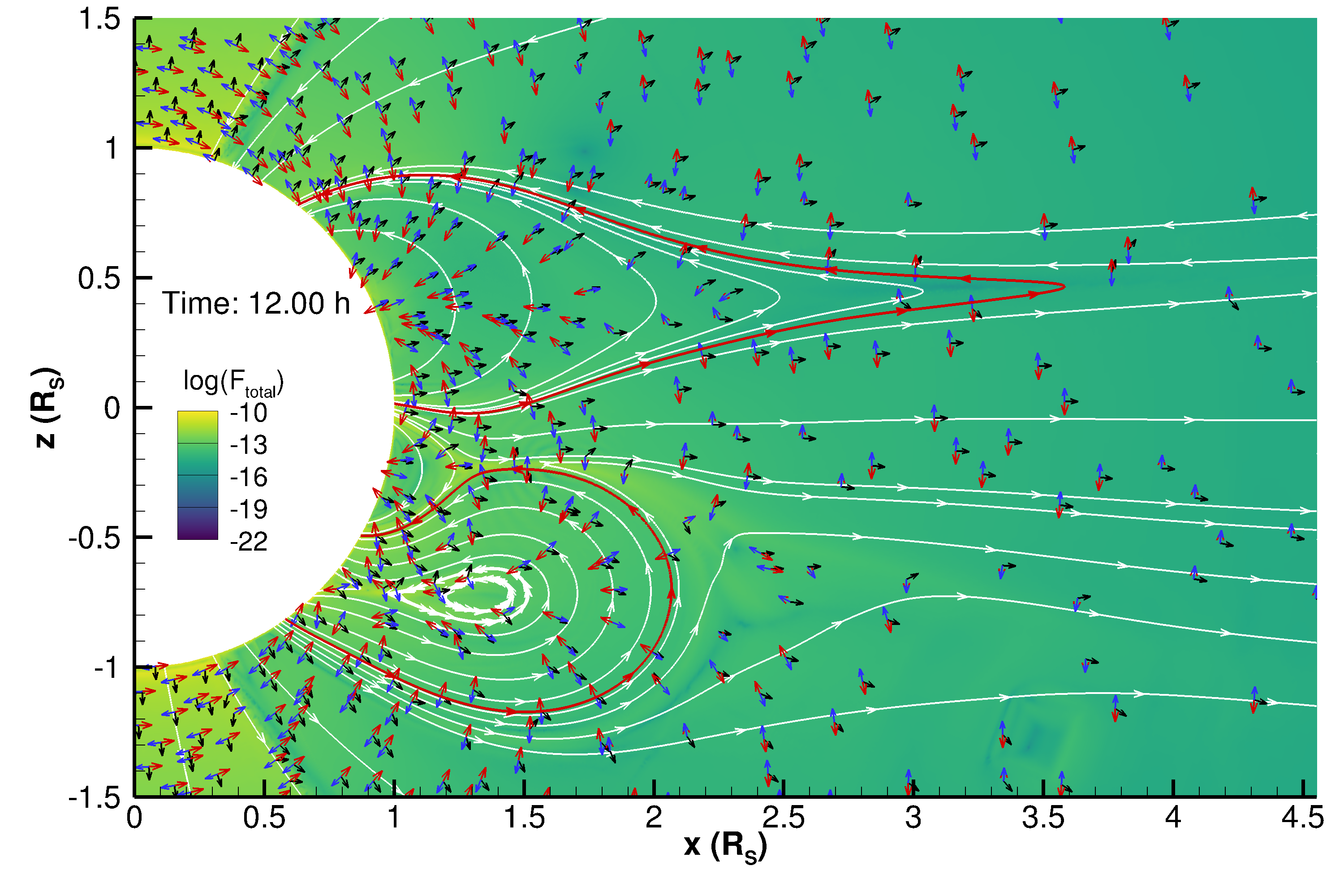}
	\caption{Simulation snapshots of the total force (colour scale) during the single eruption in the slow-wind (top panel) and faster-wind case (bottom panel). The white and red lines represent selected magnetic field lines. The vectors indicate the following forces: $\mathbf{T}_B$ (red), $-\nabla P_B$ (blue), and $-\nabla P_P$ (black). The snapshots were taken at 14 h (top) and 12 h (bottom) from the start of shearing, such that the front of both CMEs is at $\approx$ 2 R$_{\sun}$.}
	\label{fig:diff_overlying_field_lines}  
\end{figure*}

\begin{figure*}   %%% I moved this figure from the section "Formation of the stealth ejecta" so that it stays on the proper page
	\centering
	\begin{overpic}[width=0.9\textwidth]{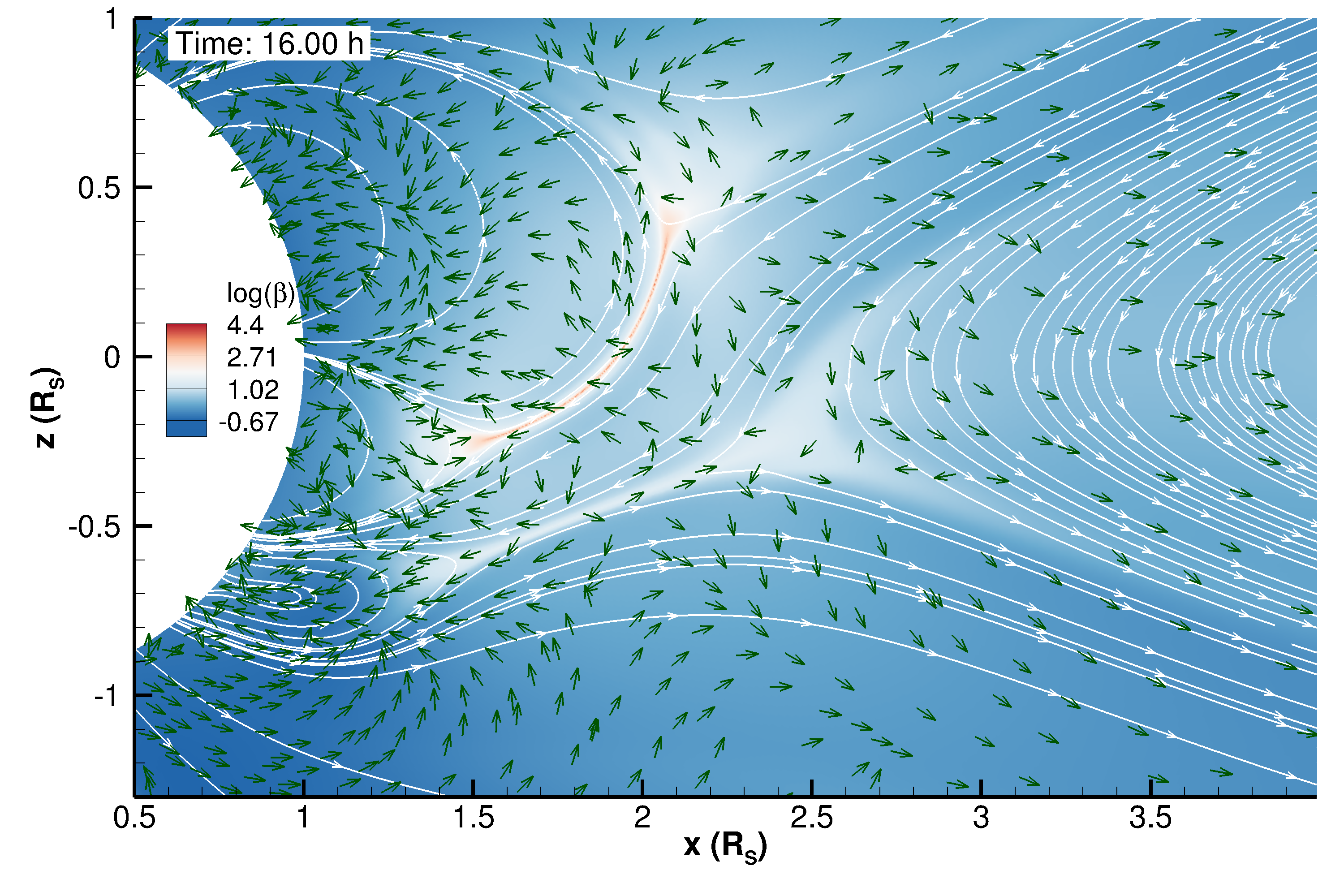}
		\put(85,40){\colorbox{white}{\parbox{0.053 \linewidth}{\textbf{CME1}}}}
		\thicklines             
		\put(22,22.5){\color{red} \circle{14}}  
	\end{overpic} 
	\caption{Simulation snapshot of the first erupting CME and its trailing current sheet, taken 16 h after the start of shearing. The colour scale depicts plasma $\beta$ values, the white lines are selected magnetic field lines, and the green vectors show the direction of the total force. The red circle highlights the second flux rope, formed from the shearing motions, that falls back to the Sun.}
	\label{fig:sw_stealth_current_sheet}    
\end{figure*}

The top panel of Fig. \ref{fig:forces_1.5Rs_winds} shows the initial difference between the two background solar wind simulations, specifically, a larger plasma pressure gradient in the faster wind case. This is caused by the approach we used to generate the faster wind by increasing the initial temperature of the simulated slow wind. The coded MHD equations governing the plasma dynamics are expressed in conservative variables, and so the temperature changes the momentum, and therefore the velocity as well as density. The left-most and right-most peaks indicated by the dashed blue lines at $\approx \pm 67 \degr$ denote compression regions between the latitudinal components of the slow and fast wind, which can also be seen as regions of enhanced electric current density in the maps of Fig. \ref{fig:peaks_identification}. The three intermediate peaks for the faster solar wind simulation (red line) lie at the borders of the northern arcade (at $\approx 35 \degr$ and $\approx 3 \degr$) and the current sheet created by the southern pseudostreamer (at $\approx -24 \degr$), also indicated by the white circles in the bottom panel of Fig. \ref{fig:peaks_identification}. These peaks clearly show the separation between the two magnetic structures. In contrast, the slow solar wind simulation (black line) only shows two peaks at the northern edge of the large streamer (at $\approx 46 \degr$) and the southern one (at $\approx -30 \degr$). These two peaks can be traced to the solar wind configuration by the white circles in the top panel of Fig. \ref{fig:peaks_identification}. The second panel of Fig. \ref{fig:forces_1.5Rs_winds} shows that there are no significant differences in $-\nabla P_B$ between the two background solar wind configurations, while the total force in the bottom panel differs mainly in the fast solar wind region due to the faster and denser plasma, from $\approx \pm 67 \degr$ poleward. On the other hand, the magnetic tension in the third panel is much stronger in the slow-wind case than in the faster-wind case. It is also more extended in latitude because the larger northern arcade connects topologically to the other two smaller arcades (Fig. \ref{fig:peaks_identification}). As discussed below, this property of the configuration has an impact on the simulated eruptions, as shown in Fig. \ref{fig:diff_overlying_field_lines}, which shows the forces that are present after the first CME is triggered by the applied shearing motions in the southern arcade. The snapshots are taken when the CME front is at $\approx$~2~R$_{\sun}$. In both the slow (top panel) and faster (bottom panel) solar wind cases, the CME is deflected towards the equatorial plane by $-\nabla P_B$ and $\mathbf{T}_B$ surrounding the arcade, indicated in Fig. \ref{fig:diff_overlying_field_lines} by the blue and red vectors, respectively. The magnetic tension is oriented towards the equator in all of the southern part of the arcade, and $-\nabla P_B$ is pushing the CME from above in approximately the same direction. The resultant of these two vectors causes the CME to expand asymmetrically towards the north. In the slow solar wind cases (top panel of Fig. \ref{fig:diff_overlying_field_lines}), this leads to magnetic reconnection between the CME flux rope and the northern arcade, creating large overlying (red) field lines. These give rise to the predominantly sunward magnetic tension (red vectors inside the red field line), which decelerate the eruption. In the faster solar wind case, the bottom panel of Fig. \ref{fig:diff_overlying_field_lines} clearly shows the magnetic separation between the CME and the northern streamer, as indicated by the red lines, which does not enclose the flux rope, enabling it to erupt more easily. This also facilitates the expansion of the southern arcade, explaining the quicker formation of the flux rope in the faster solar wind cases.

\subsection{Formation of the stealth ejecta} \label{subsec:formation of stealth}

\begin{figure*}
        \centering
        \includegraphics[width=0.9\textwidth]{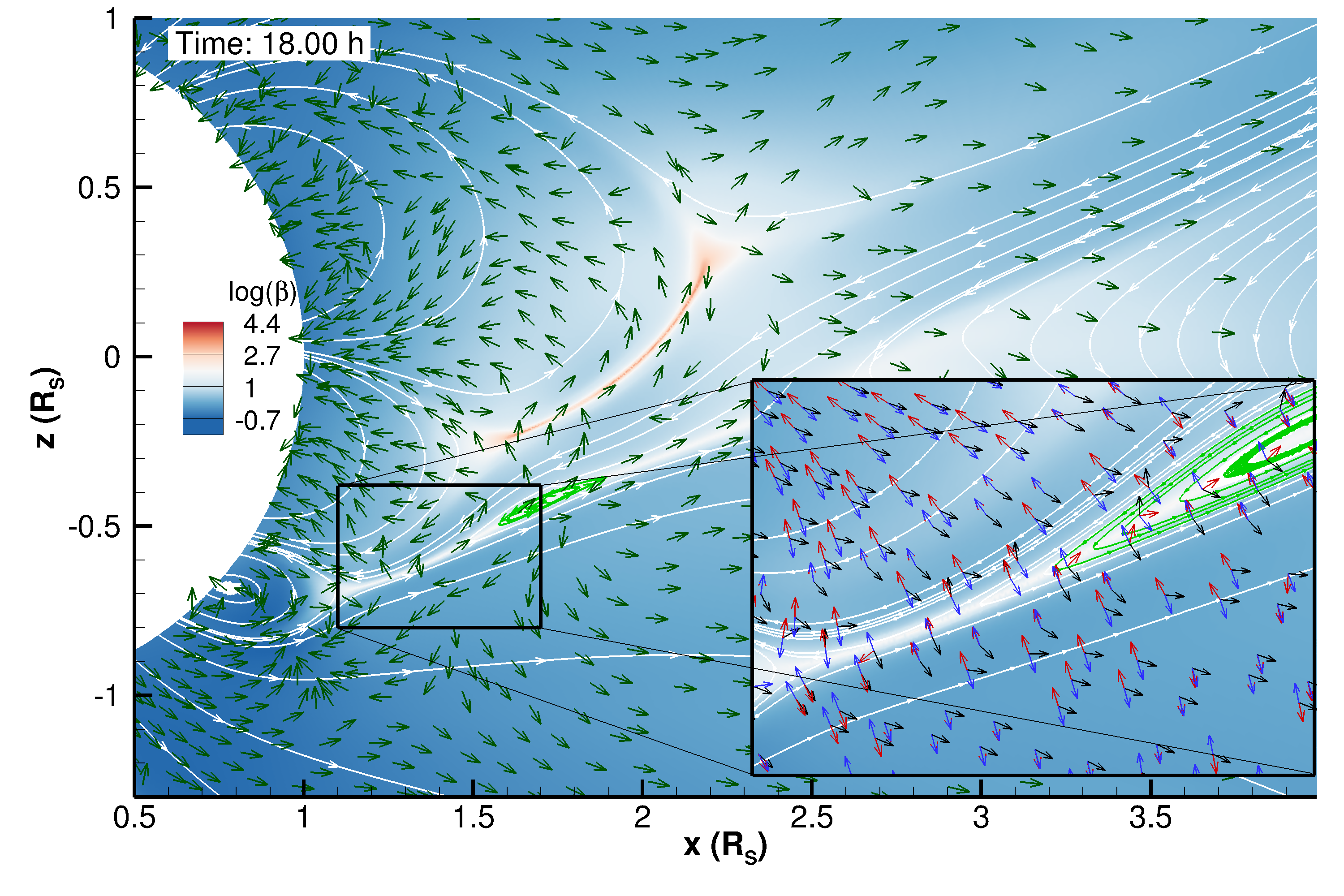}
        \vspace{-10px}
        \caption{Same as Fig. \ref{fig:sw_stealth_current_sheet}, but for a larger field of view, taken 18 h after the start of shearing. The green field lines represent the stealth ejecta formed in the current sheet trailing CME1. Bottom right panel: Enlarged view of the black rectangle showing the current sheet in which the stealth ejecta forms; the colour scale depicts plasma $\beta$, and the vectors are $\mathbf{T}_B$ (red), $-\nabla P_B$ (blue), and $-\nabla P_P$ (black).}
        \label{fig:sw_stealth_formed_flux_rope} 
\end{figure*}

In this subsection, we explain the formation of the stealth ejecta in the slow-wind case (eruption + stealth scenario). For this purpose, we calculated and analysed the forces in the region of the current sheet trailing the first CME (CME1) in Fig. \ref{fig:sw_stealth_current_sheet}, which shows plasma $\beta$, magnetic field lines, and total force vectors at 16 h after the start of shearing. The important elements in this scenario are the first CME triggered by the shearing motions, one flux rope formed through the same mechanism, and another flux rope formed by the reconfiguration of the coronal magnetic field. Consistent with the sunward total force vectors (green), the second flux rope, indicated in Fig. \ref{fig:sw_stealth_current_sheet} by the red circle, falls back to the Sun. The total force vectors also determine the motion of CME1 during its eruption away from the Sun. In this process, two current sheets are formed, one trailing CME1 extending to the second flux rope, and the second between the CME and the compressed northern arcade. Both are clearly depicted in Fig. \ref{fig:sw_stealth_current_sheet} by the high plasma $\beta$ regions and the anti-parallel magnetic field lines. The two flux ropes moving in opposite directions stretch the current sheet trailing CME1 and create a region in which the total force vectors point in opposite directions at each end of the current sheet.\\
Figure \ref{fig:sw_stealth_formed_flux_rope} shows how the simulation evolved 18 h after the shearing commences. The $-\nabla P_B$ blue vectors shown in Fig. \ref{fig:sw_stealth_formed_flux_rope} indicate compression of the sides of the current sheet, creating a flux rope highlighted by the green magnetic field lines, which is seen more clearly in the expanded view. Since the height at which the reconnection sets in to form this structure is $\approx$ 1.4 R$_{\sun}$, high enough to present no clear low-coronal signatures \citep{robbrecht_stealth}, we suggest that this is a stealth ejecta. This blob-like ejecta follows the total force vectors and erupts along a similar path as CME1.

\subsection{Formation of plasma blobs} \label{subsec:formation of plasma blobs}

In this subsection we focus on another type of structure, plasma blobs. They are formed only in the slow solar wind cases by magnetic reconnection in the equatorial current sheet as it rebuilds in the aftermath of the eruptions. For simplicity, the scenario we analysed is that of the single eruption, but the physical mechanisms and explanations are valid for all three slow-wind simulations.\par 
As discussed in Sec. \ref{subsec:simulations}, all the simulated CMEs experience deflection towards the equatorial plane, but only those ejected into the slow wind magnetically reconnect with the northern arcade during their eruption phase, becoming structurally coupled to it, as in the top panel of Fig. \ref{fig:diff_overlying_field_lines}. In this process, the CMEs compress the arcade, which after the passage of the CME slowly returns to its original size.\par
\begin{figure}[h!]
        \centering
        \includegraphics[width=0.5\textwidth, trim={2cm 0 1cm 1.5cm}, clip]{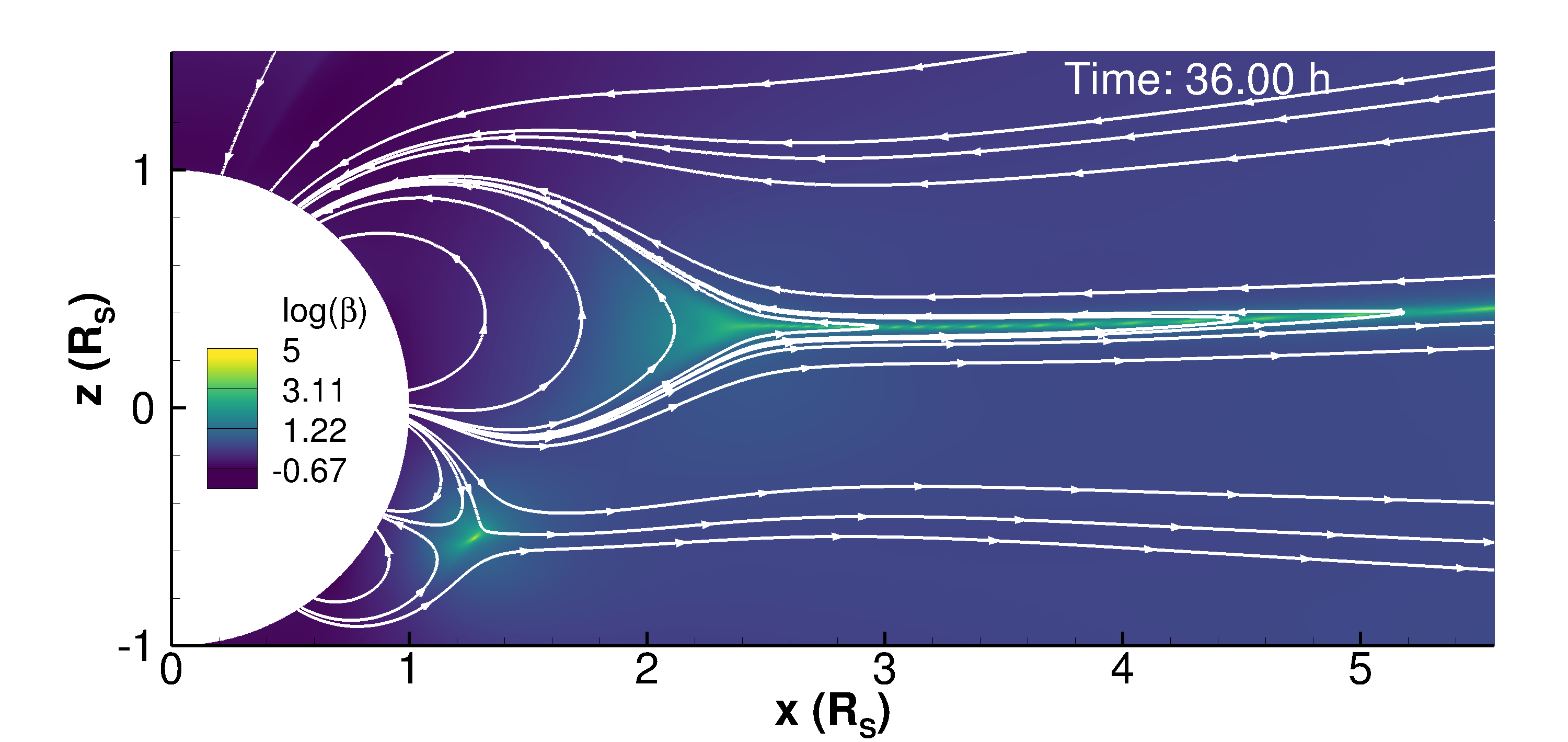}\\
		\begin{overpic}[width=0.5\textwidth, trim={2cm 0 1cm 1.5cm}, clip]{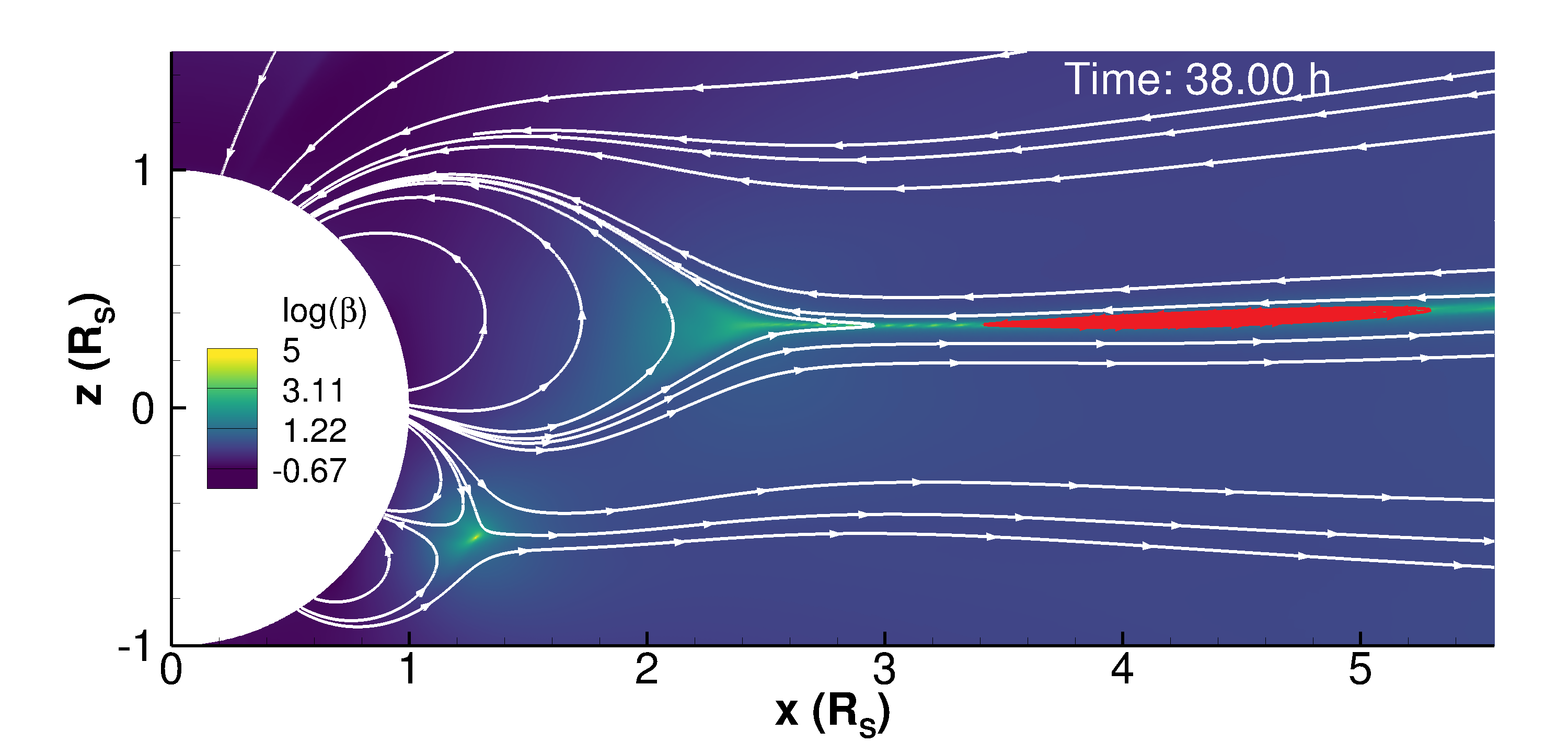}
			\put(56,27){\tiny \textbf{x}}
		\end{overpic}
        \caption{Snapshots of the plasma $\beta$ from the slow wind - single eruption simulations taken at 36 h (top panel) and at 38 h (bottom panel), depicting the formation of a plasma blob (red field lines) from the northern streamer cusp. The X-point created between the blob and the streamer cusp is indicated (X) in the bottom panel.}
        \label{fig:blob_formation}
%       \vspace{-15px}  
\end{figure}
\begin{figure}[h!]
        \centering
        \includegraphics[width=0.5\textwidth]{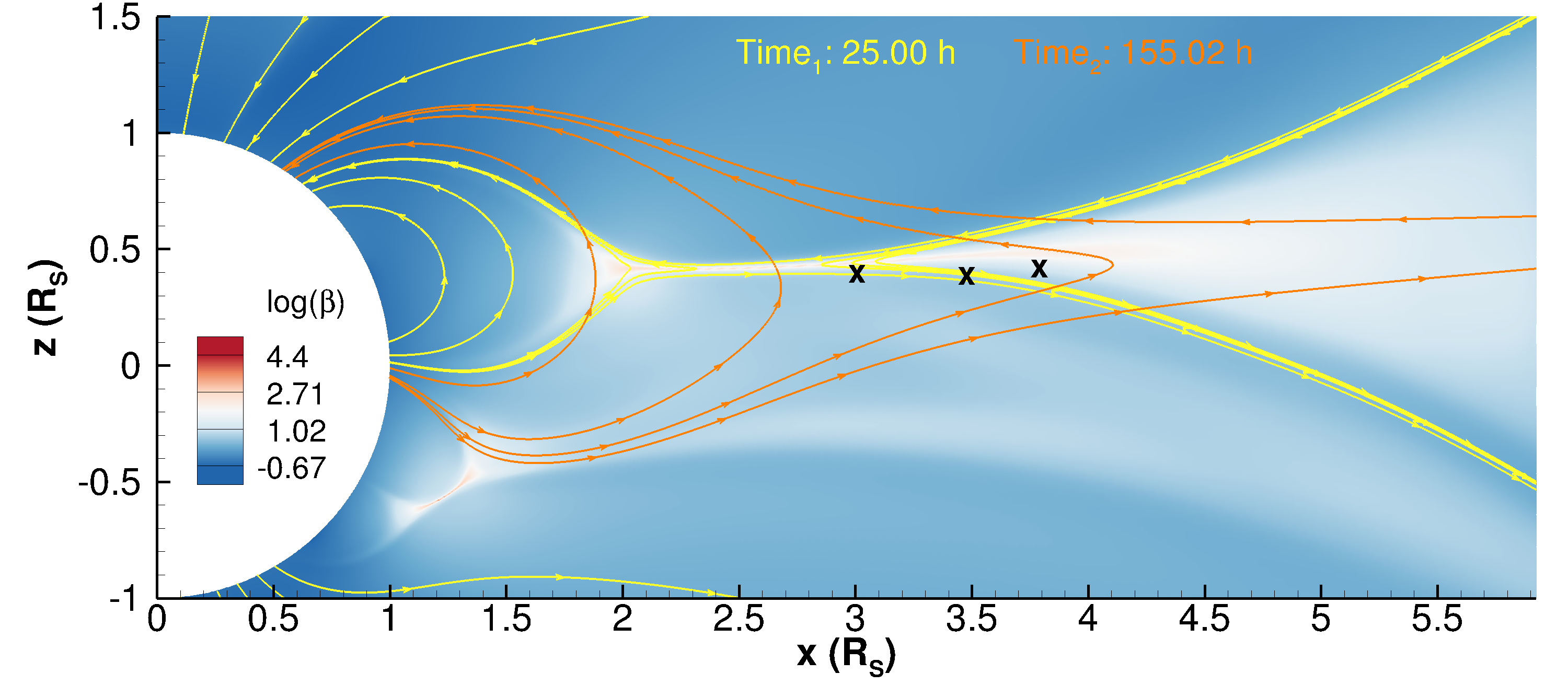}
        \vspace{-5px}
        \caption{Snapshot of the plasma $\beta$ (colour scale) and magnetic field lines (yellow) from the slow wind - single eruption simulation taken 25 h from the start of shearing, showing the compressed arcade and the trailing part of the CME on the right side. The orange field lines show the magnetic configuration of the northern arcade from the same simulation, after it has relaxed to its original state, 155 h from the start of shearing. The X-points form between the streamer cusp and blobs and evolve further from the Sun as time increases.}
        \label{fig:blobs_x_points}
        \vspace{-15px}  
\end{figure}

Here we address how the blobs are formed and why they are created. After the CME has erupted into the equatorial current sheet, the cusp of the northern arcade elongates (top panel of Fig. \ref{fig:blob_formation}) and pinches off at an X-point indicated in the bottom panel of Fig. \ref{fig:blob_formation}, creating the plasma blob depicted by the red field lines. This mechanism resembles the one described by \citet{wang_blob_formation} in their Fig. 8b. In their study, they called such a structure a detached `plasmoid', although they deemed this scenario unlikely in a realistic 3D geometry in which a helical flux rope still attached to the Sun would form instead. However, this scenario is not entirely impossible since the coronal structure in which they placed this mechanism was a stable streamer, in contrast to our simulations, where the blobs form from the cusp of an expanding streamer. In order to illustrate this dynamic property, in Fig. \ref{fig:blobs_x_points} we superpose the orange magnetic field lines of the relaxed streamer 155 h from the start of shearing on those of the perturbed state (yellow field lines) only 25 h from the start of shearing, where the CME is still visible in the right panel of the figure. The recovery of the small compressed arcade to its original size can be tracked via the location of the X-points created between the blobs and the streamer cusp, indicated in Fig. \ref{fig:blobs_x_points}. The rising streamer cusp is renewed every time a blob detaches. As the arcade evolves, the X-points are placed at higher distances from the Sun, indicating the expansion of the magnetic structure. It would be interesting to investigate this scenario further in a 3D simulation without the limitations of the 2.5D geometry used here.\par

\begin{figure}[h!]
        \centering
        \includegraphics[width=0.5\textwidth]{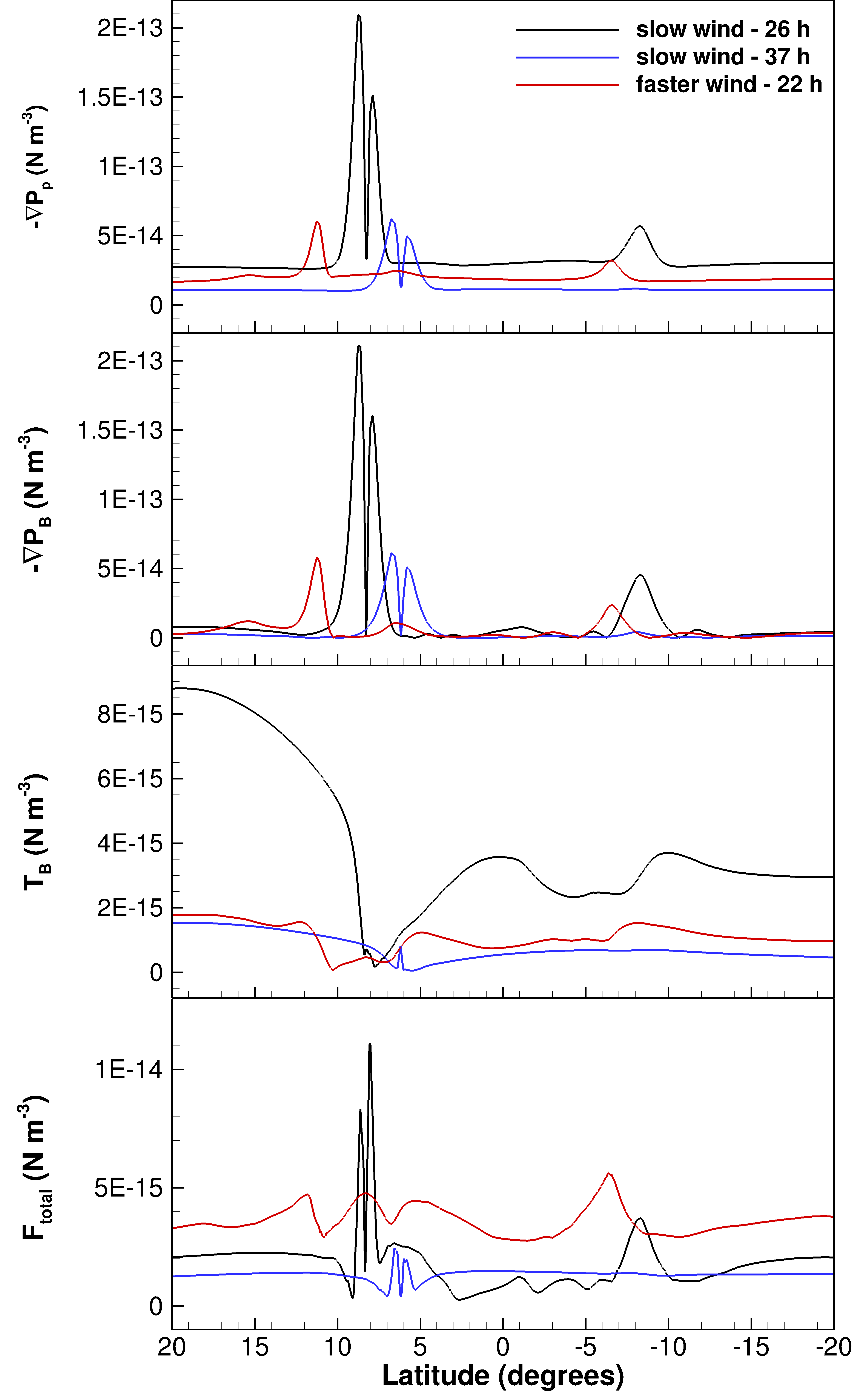}
        \vspace{-10px}
        \caption{Forces extracted from a latitudinal cut through the current sheet occurring in the aftermath of eruptions at different heights and simulation times. The slices were taken through the slow wind - single eruption simulation at 2.62 R$_{\sun}$ and 26 h from the start of shearing (black line), and at 3.2 R$_{\sun}$ and 37 h (blue line), and through the faster wind - single eruption at 3.3 R$_{\sun}$, 22 h from the start of shearing (red line).}
        \label{fig:forces_CS_after_eruption}
        %       \vspace{-15px}  
\end{figure}

\begin{figure*}
	\centering
	\includegraphics[width=1\textwidth]{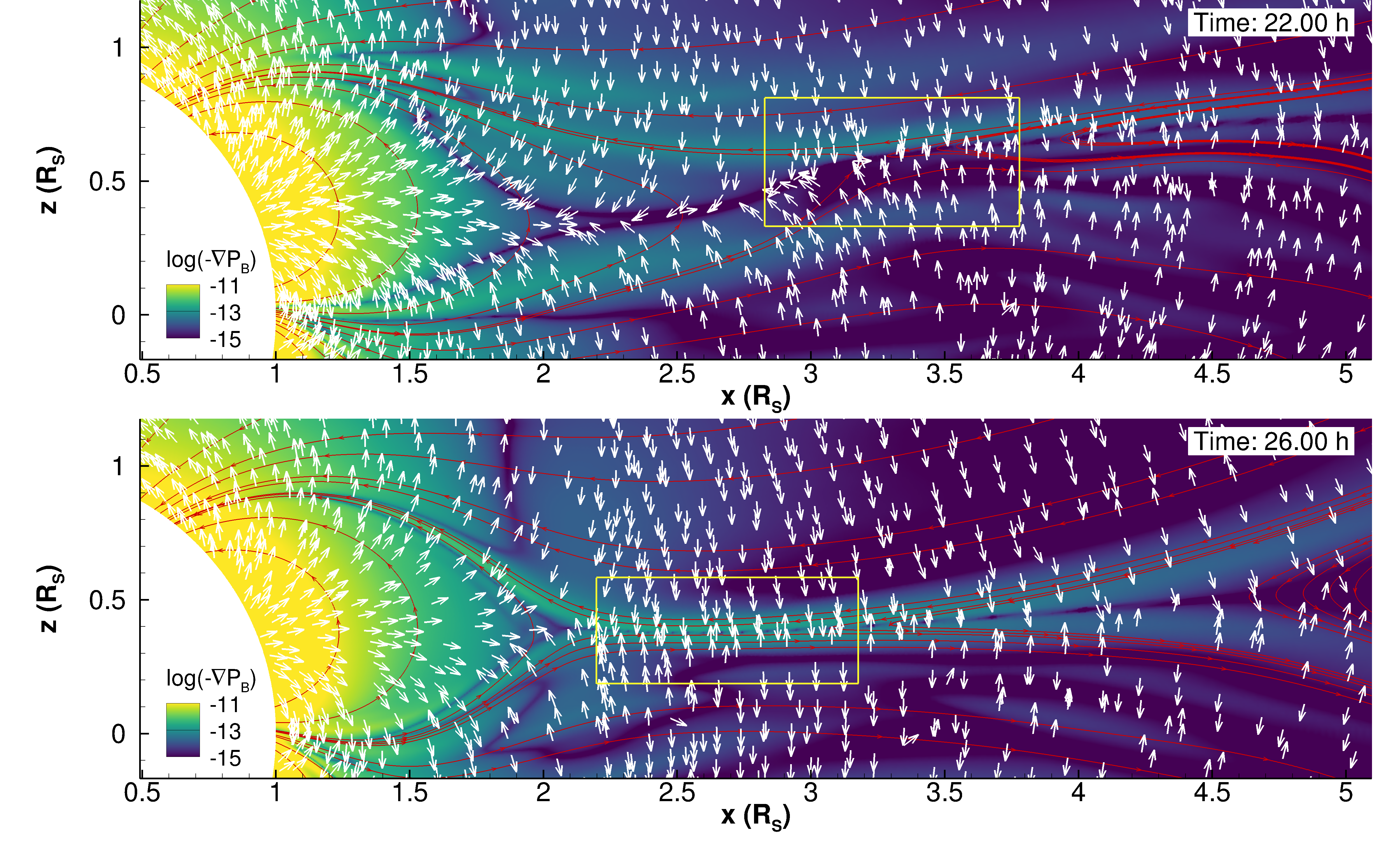}
	\vspace{-10px}
	\caption{Snapshots of the magnetic pressure gradient (colour scale) depicting the formation of current sheets in the aftermath of CMEs in the single-eruption case inserted into the faster background wind (top panel) and slow wind (bottom panel). The white vectors represent $-\nabla P_B$, and the red lines are selected magnetic field lines. The yellow rectangles are regions of interest, as discussed in the text.}
	\label{fig:2D_image_current_sheets}
	%       \vspace{-15px}  
\end{figure*}

We calculated the forces in the region trailing the CME, where these blobs are formed. In particular, we extracted their values along a latitudinal arc at constant distance from the Sun in order to study the forces in the direction perpendicular to the current sheet that led to magnetic reconnection. For the slow-wind simulation, we took these values just before the formation of two of the plasma blobs, in the vicinity of the reconnection site (close to the X-points indicated in Fig. \ref{fig:blobs_x_points}) 26 h and 37 h after the start of shearing. Since the CMEs in the faster-wind cases did not present such blobs, we also took a cut through the current sheet trailing the single eruption in this latter scenario in order to understand the differences that lead to the formation of the small flux ropes. The values of the extracted forces are plotted as a function of latitude in Fig. \ref{fig:forces_CS_after_eruption}. The first noticeable distinction is between the profiles of $-\nabla P_P$ and $-\nabla P_B$ for the two background solar winds. The current sheet in the aftermath of the single eruption in the slow wind is associated with double-peaked structures in both these parameters, as compared to the single peaks associated with the current sheet in the faster-wind case. Even though the amplitudes of these peaks decrease in time from the formation of one blob to the other (back line versus blue line) due to the strong disturbances being carried away from the Sun, the double-peaked profile is preserved. We interpret the double peaks as the cause of the formation of the blobs, since the current sheet is squeezed from the sides and magnetic reconnection is induced by the perpendicular-oriented $-\nabla P_B$ vectors, which are shown in the bottom panel of Fig. \ref{fig:2D_image_current_sheets} (inside the yellow rectangle). Figure \ref{fig:2D_image_current_sheets} shows log$(-\nabla P_B)$ values in the current sheet formed in the aftermath of the CME in the faster (top panel) and slow (bottom panel) solar wind cases. The squeezing of the current sheet does not occur in the faster-wind case because the plasma and magnetic pressure gradients are strong only on the northern side of the current sheet, preventing the magnetic field from pinching. This may be visualised in the yellow rectangle in the top panel of Fig. \ref{fig:2D_image_current_sheets}. The third panel of Fig. \ref{fig:forces_CS_after_eruption} depicts the magnetic tension. This is almost two orders of magnitude smaller than the other forces, and therefore the Lorentz force is mainly comprised of the magnetic pressure gradient. This is not surprising since the current sheet contains magnetic field lines that are nearly parallel to each other, so the lack of curvature leads to extremely low values of $\mathbf{T}_B$.\par

Other plasma blobs are present in the same current sheet as well much later, after the eruption of the CME. They are not from the streamer cusp, and they occur at larger distances from the Sun. In the slow wind - double eruption case, the second CME catches up with the first CME and compresses it, creating a current sheet between the two flux ropes where plasma blobs also form. All these features are driven by interesting mechanisms, but their analysis is not addressed in this paper.

\subsection{Propagation and interaction of CMEs} \label{subsec:propagation CMEs}

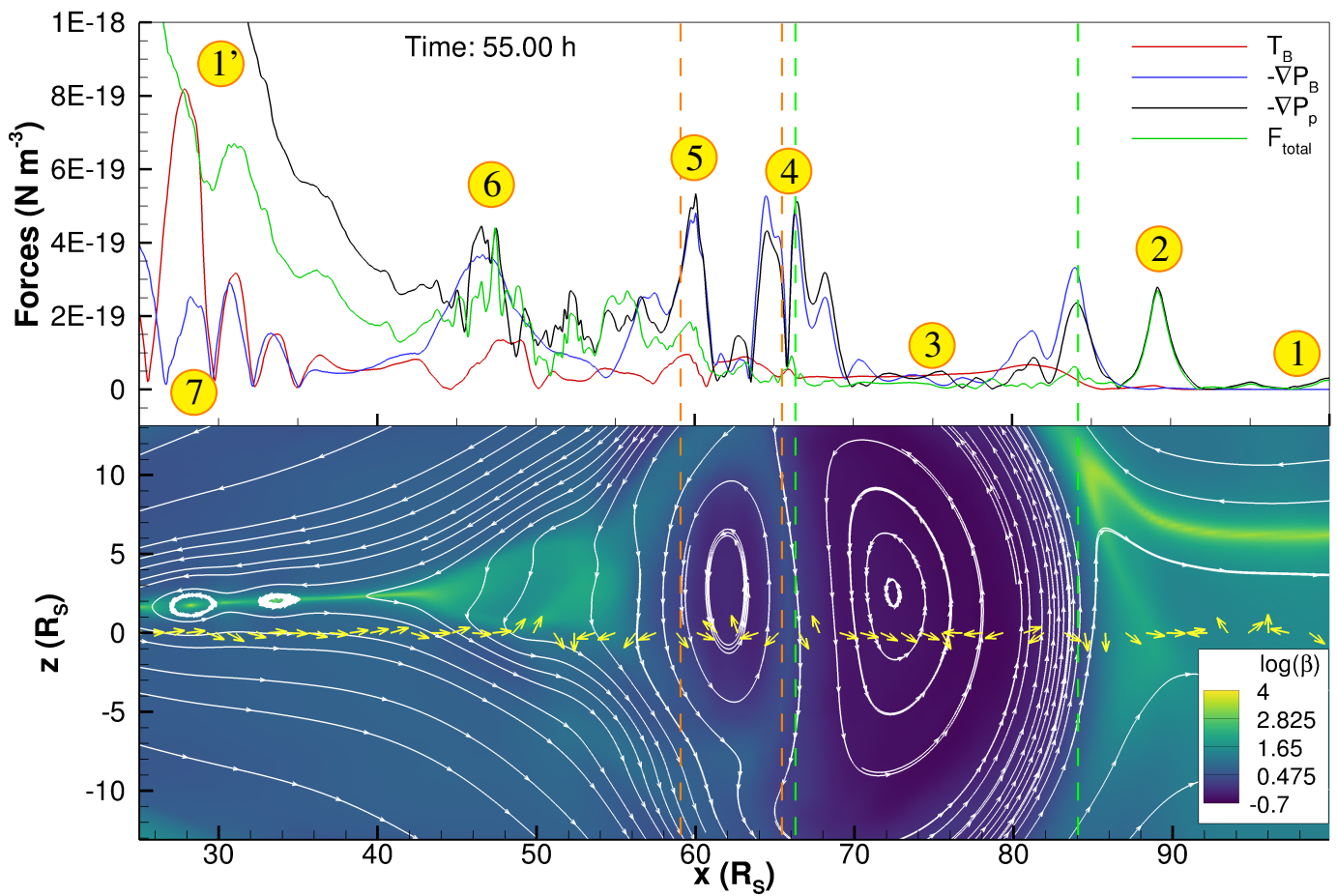
\begin{figure*}
        \centering
        \begin{overpic}[width=1\textwidth]{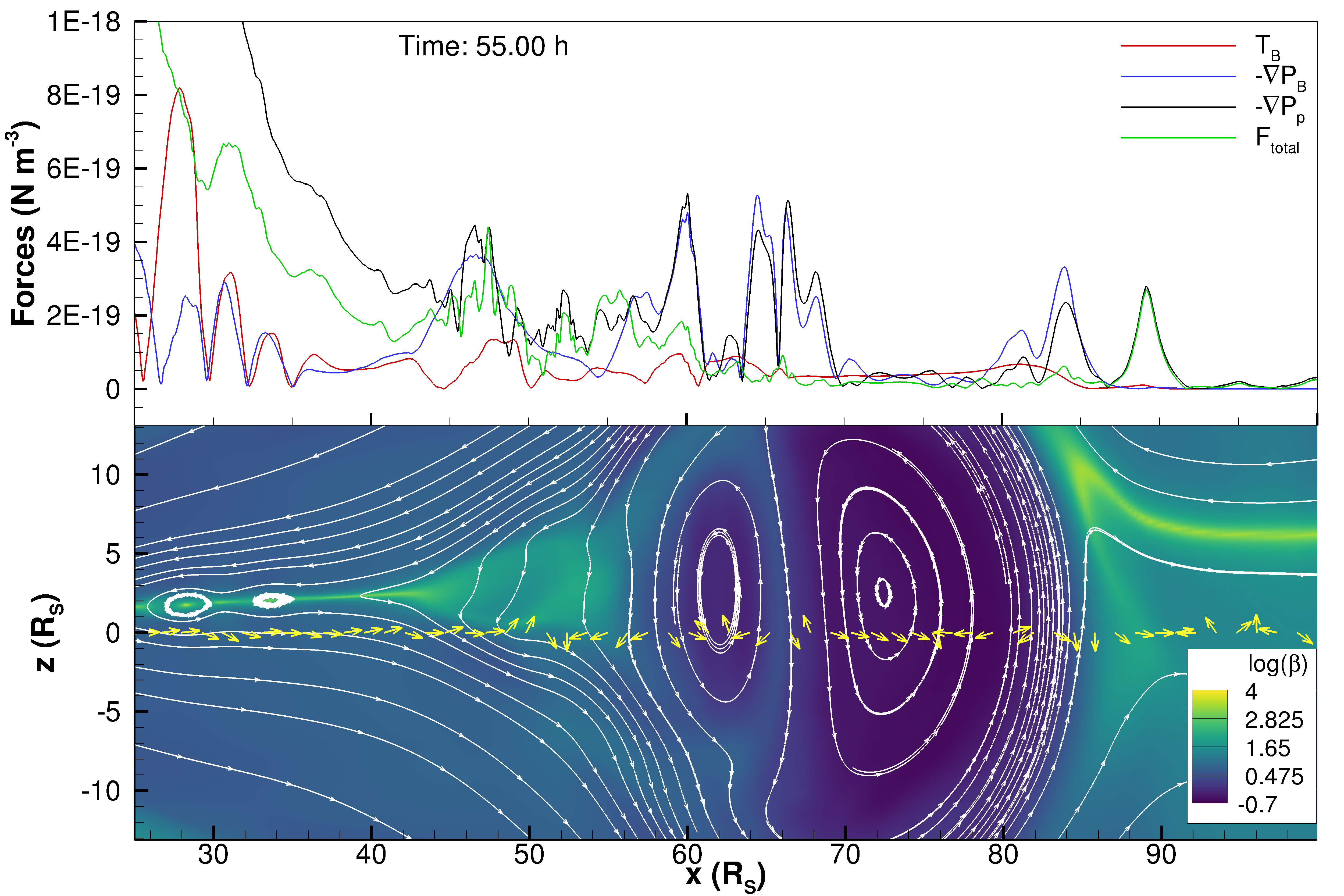}
                \put(95.6,40.4){\color{orange} \circle*{3.5}}
                \put(95.6,40.4){\color{yellow} \circle*{3.2}}
                \put(95,39.5){\Large \text{1}}

                \put(85.4,48.2){\color{orange} \circle*{3.5}}
                \put(85.4,48.2){\color{yellow} \circle*{3.2}}
                \put(84.8,47.3){\Large \text{2}}

                \put(0,0){\color{green} \thicklines \dashline[30]{1.5}(79.4,4.5)(79.4,65)}
                \put(68.6,41){\color{orange} \circle*{3.5}}
                \put(68.6,41){\color{yellow} \circle*{3.2}}
                \put(68,40.2){\Large \text{3}}
                \put(0,0){\color{green} \thicklines \dashline[30]{1.5}(58.5,4.5)(58.5,65)}

                \put(0,0){\color{orange} \thicklines \dashline[30]{1.5}(57.5,4.5)(57.5,65)}
                \put(0,0){\color{orange} \thicklines \dashline[30]{1.5}(50,4.5)(50,65)}
                \put(58,54){\color{orange} \circle*{3.5}}
                \put(58,54){\color{yellow} \circle*{3.2}}
                \put(57.4,53.1){\Large \text{4}}

                \put(51,55){\color{orange} \circle*{3.5}}
                \put(51,55){\color{yellow} \circle*{3.2}}
                \put(50.4,54.1){\Large \text{5}}

                \put(36,53){\color{orange} \circle*{3.5}}
                \put(36,53){\color{yellow} \circle*{3.2}}
                \put(35.4,52.1){\Large \text{6}}

                \put(14,37.5){\color{orange} \circle*{3.5}}
                \put(14,37.5){\color{yellow} \circle*{3.2}}
                \put(13.4,36.6){\Large \text{7}}

                \put(16,62){\color{orange} \circle*{3.5}}
                \put(16,62){\color{yellow} \circle*{3.2}}
                \put(15.2,61.1){\Large \text{1'}}
        \end{overpic}        
        \caption{Snapshots of the evolution of the CMEs in the slow wind - double eruption case, taken 55 h after the start of shearing. Top: Forces extracted from an equatorial 1D slice through this configuration. Bottom: Snapshot of the plasma $\beta$ (colour scale) and selected magnetic field lines. The 1D slice in the top panel was extracted from this same configuration. The numbers indicate the following features, described more extensively in the text: 1 and 1' - solar wind in front of or trailing the CMEs; 2 - plasma compression front; 3 - core of CME1; 4 - compression zone between the two CMEs; 5 - end of second flux rope; 6 - tail of the CMEs; and 7 - plasma blobs. The dashed green and orange lines delimit the two flux ropes. The yellow vectors represent the total force along the 1D slice.}
        \label{fig:forces_double_eruption}      
\end{figure*}

In this final subsection, we analyse the interaction between CMEs and the interaction with the background solar wind. We discuss the first topic in the context of the slow wind - double eruption simulation because it contains two CMEs triggered by the imposed shearing motions, and they are large enough to influence each other during their propagation to 1 AU. We calculated the forces in the entire computational domain, and then extracted the values along the equator, which is shown in the top panel of Fig. \ref{fig:forces_double_eruption}. The time of the simulation chosen to perform these steps is 55 h after the start of shearing, when the front of the first CME is at approximately 85 R$_{\sun}$. At this distance from the Sun, the second CME has not yet merged with the first, but it is close enough to influence it, as is shown in the bottom panel of Fig. \ref{fig:forces_double_eruption}. Regions indicated by numbers 1 and 1' depict the background solar wind in front of and behind the eruptions, where plasma $\beta$ is much higher than 1 and $-\nabla P_P$ (black line) makes the dominant contribution to the total force (green line). The peak indicated by number 2 is present only in $-\nabla P_P$ and is created by the solar wind plasma accumulation in front of the CME, as it compresses the material ahead of it while propagating through interplanetary space. The first flux rope (number 3) is delimited by two peaks in $-\nabla P_P$ and $-\nabla P_B$ and by the dashed green lines, while inside, all the forces plateau at low values. The yellow vectors show that the total force is directed oppositely between the front and rear of the CME interior, flattening the flux rope and creating the well-known `pancake' effect. This is caused by the radial expansion of the solar wind and occurs regardless of the CME-CME interaction. We just mention here that this realistic effect is also present in our simulations, but we do not further expand on the topic since it has been extensively studied observationally and by numerical means by many authors \citep[e.g.][]{riley_crooker_2004, savani_2011, isavnin_2016}. The vectors do not change orientation exactly at the centre of the flux rope, indicating that the centre will be carried towards the front part of the CME. Number 4 depicts the zone of compression between the two CMEs, and also the beginning of the second flux rope, delimited by the dashed orange lines. The behaviour of the total force vectors within the second CME is very similar to that of the first CME, pushing the front and rear towards the centre and producing the same pancaking effect. As the two flux ropes approach each other, they magnetically reconnect, and the second flux rope disappears as it fully merges with the first during their propagation. This second process is able to take place because the two flux rope share the same magnetic orientation, as they originate from the same coronal arcade. This leads to the oppositely directed fields at the trailing edge of the first flux rope and the leading edge of the second flux rope. The overlying magnetic field lines that surround both CMEs are evidence that this merging has already started to take place. The region between the end of the second flux rope (number 5) and the peak at number 6 represents the tail area of the CMEs, where almost all the forces are turbulent and have high values. The magnetic tension does not contribute greatly to the propagation of the eruptions, as compared to the initiation and eruption phases. However, $\mathbf{T}_B$ and $-\nabla P_B$ do show major fluctuations in the region trailing the CMEs. The peaks in the area depicted by number 7 are the cause of magnetic null points either at the centre of or between plasma blobs created in the aftermath of the eruptions from the reconfiguration of the current sheet. We note that the 1D slice does not pass directly through the centre of the flux ropes or plasma blobs, since they propagate into the equatorial current sheet, which is deflected northward due to the magnetic configuration at the inner boundary.\par               
We now discuss the interaction of the CMEs with the background wind in the cases of the single eruptions inserted into both background solar winds. Because these eruptions are the simplest cases, we can therefore assess the influence of the solar wind alone, excluding effects of CME interaction. The shearing speed applied to obtain these single eruptions is approximately the same for both solar winds, as is their morphology and eruption mechanism. The difference, which we cannot eliminate from the contributing factors, is the presence of plasma blobs trailing the CME in the slow-wind case.\par 

\begin{figure}
	\centering
	\includegraphics[width=0.5\textwidth]{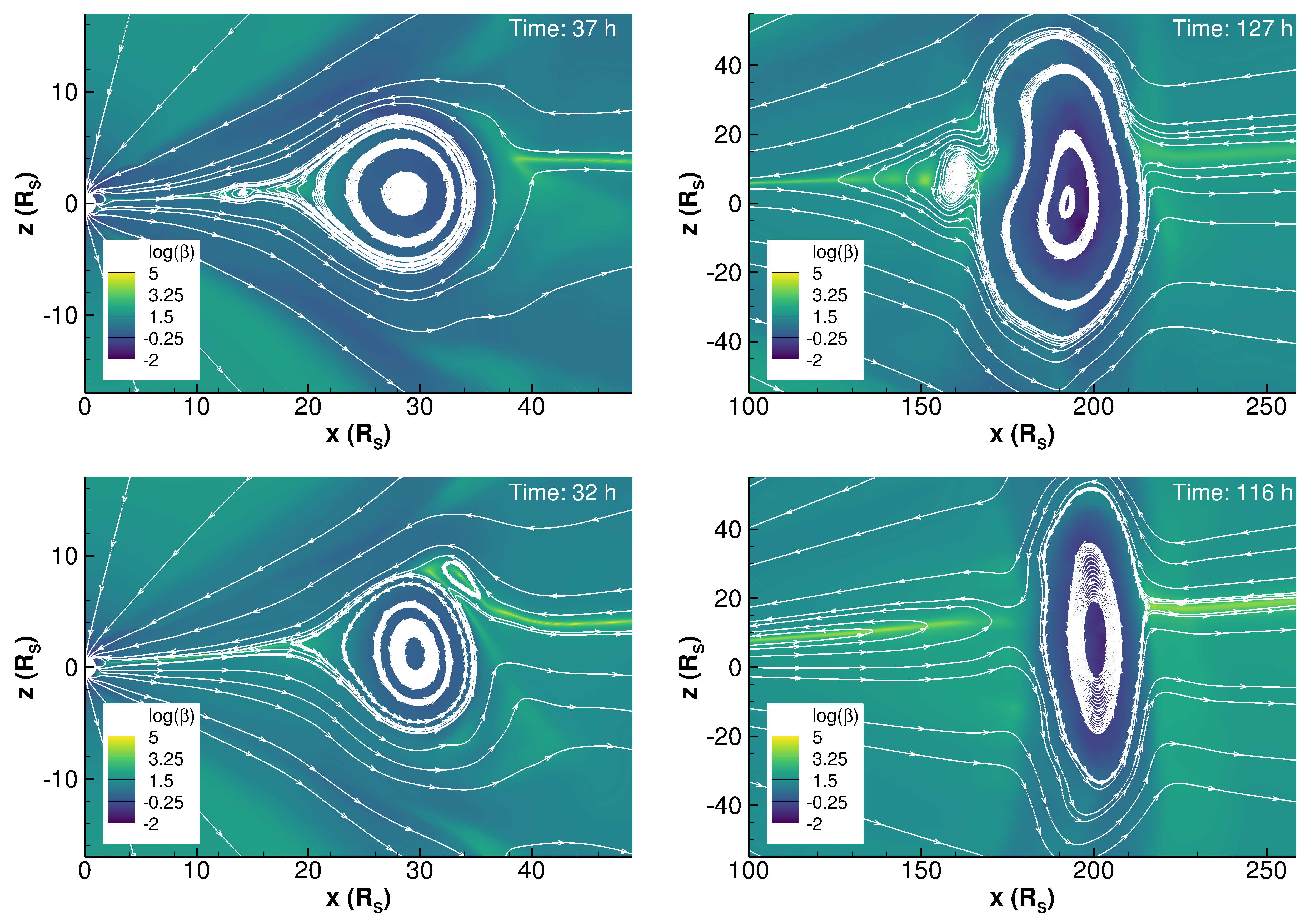}
	\caption{Snapshots of the CMEs in the single eruption cases, propagating in the slow (top row) and faster (bottom row) background solar winds. The first column shows the CMEs when the front of the flux ropes are at $\approx$35 R$_{\sun}$, and the second column when they reach 1 AU. Each snapshot was taken at the times indicated in the respective panels, counted from the start of shearing. The colour scale depicts plasma $\beta$ values, and the white lines represent selected magnetic field lines.}
	\label{fig:cmes_snapshots}
	%       \vspace{-15px}  
\end{figure}

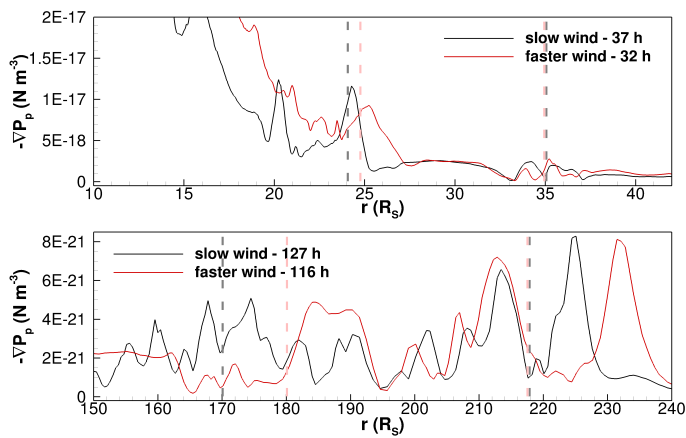
\begin{figure}
	\centering
	\begin{overpic}[width=0.5\textwidth, trim={0 0 12cm 0}, clip]{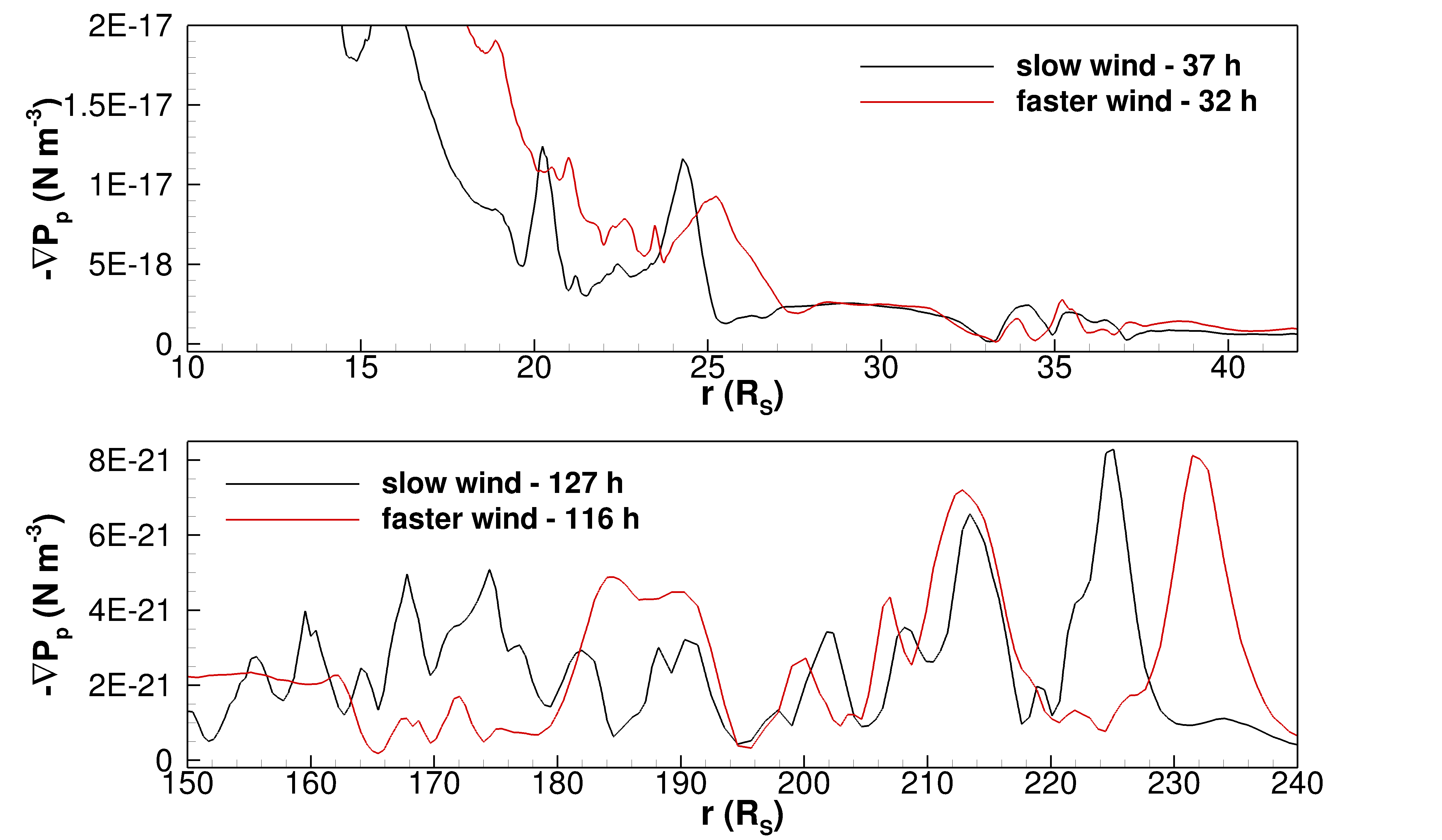}
		\put(0,0){\color{gray} \thicklines \dashline[30]{1.5}(78.6,37)(78.6,60)}
		\put(0,0){\color{gray} \thicklines \dashline[30]{1.5}(50.2,37)(50.2,60)}
		\put(0,0){\color{pink} \thicklines \dashline[30]{1.5}(78.3,37)(78.3,60)}
		\put(0,0){\color{pink} \thicklines \dashline[30]{1.5}(52,37)(52,60)}    
		
		\put(0,0){\color{gray} \thicklines \dashline[30]{1.5}(76.2,6)(76.2,29)}
		\put(0,0){\color{gray} \thicklines \dashline[30]{1.5}(32.3,6)(32.3,29)}
		\put(0,0){\color{pink} \thicklines \dashline[30]{1.5}(75.9,6)(75.9,29)}
		\put(0,0){\color{pink} \thicklines \dashline[30]{1.5}(41.5,6)(41.5,29)}
	\end{overpic}        
	\caption{Plasma pressure gradient values extracted from the equatorial cross-section through the single eruptions depicted in Fig. \ref{fig:cmes_snapshots}, propagating in the slow background solar wind (black line) and in the faster wind (red line). The panels show results for two distance ranges that include the main flux rope, which is approximately delineated by the dashed grey and pink lines in the slow and faster solar wind, respectively.}
	\label{fig:plasma_press_grad_single_er}
	%       \vspace{-15px}  
\end{figure}

\begin{figure}
	\centering
	\includegraphics[width=0.5\textwidth]{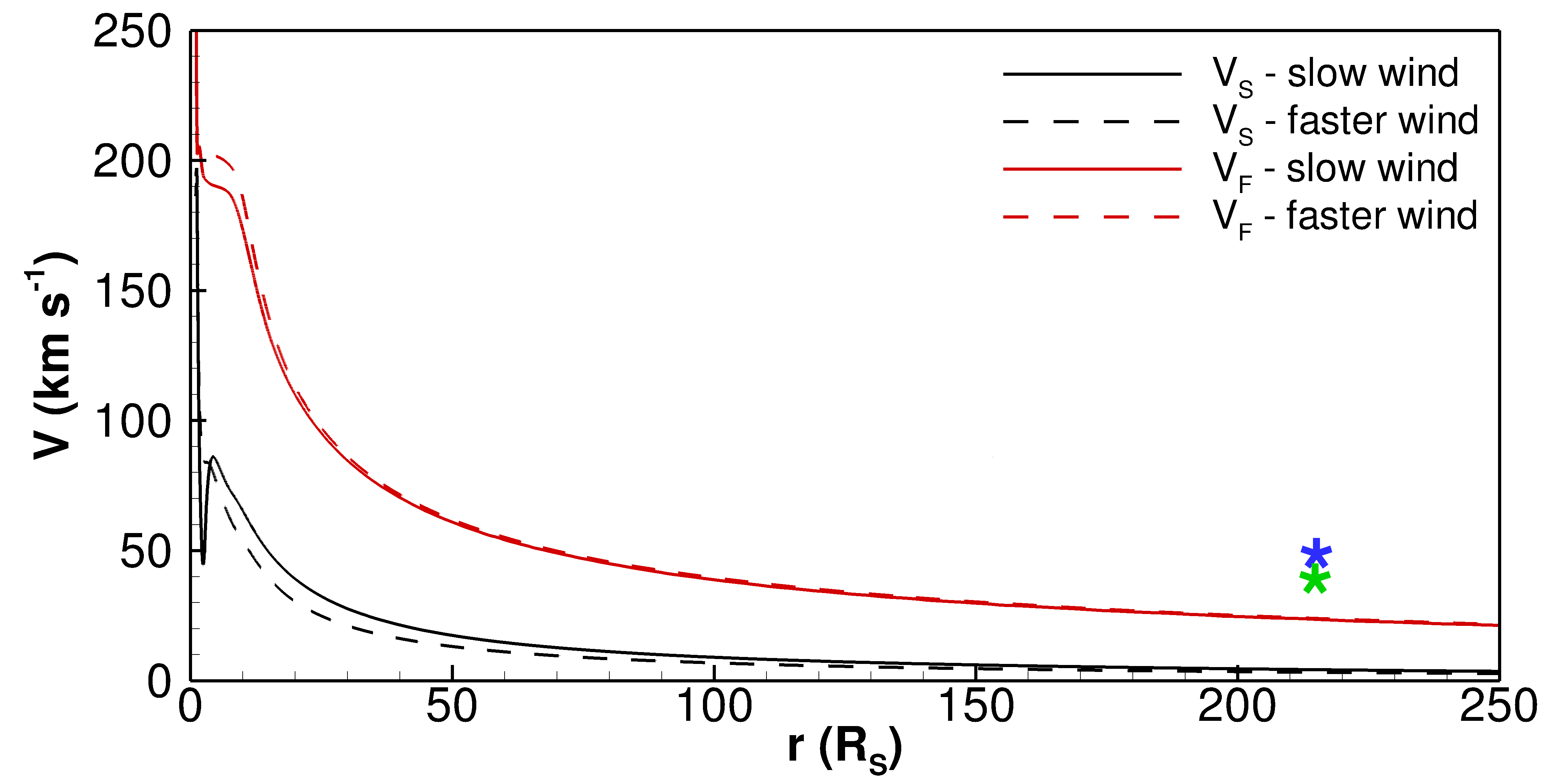}
	\caption{MHD wave velocities corresponding to slow (red lines) and fast (black lines) magnetoacoustic waves for the slow (solid lines) and faster (dashed lines) background solar winds. The velocity of the front of the CME at 1AU with respect to the solar wind is indicated by the blue asterisk in the case of the slow background wind and by the green asterisk for the faster wind.}
	\label{fig:wave_speeds}
	%       \vspace{-15px}  
\end{figure}

For this part, we performed a similar analysis as in the previously discussed double eruption scenario, but focused on the differences occurring over time due to the effects during propagation. Therefore, we analyse the forces along equatorial cross-sections through the CMEs when the fronts of their flux ropes are at $\approx$35 R$_{\sun}$ ($\approx$0.163 AU) and 1 AU. Figure \ref{fig:cmes_snapshots} shows an overview of the magnetic field configuration of the eruptions at the moment of computation of the forces. The times of the snapshots differ between the two background solar winds because of the higher propagation speed of the CME in the faster-wind case. Figure \ref{fig:plasma_press_grad_single_er} depicts $-\nabla P_P$ values taken at the equator. The top panel shows that close to the Sun, the eruptions do not differ greatly between the two background winds. The exceptions are the higher values in the faster-wind case below 20 R$_{\sun}$ (as expected due to the overall higher density), and the stronger peak in the slow-wind case, indicating the trailing edge of the flux rope. To identify the  forces in this region more easily, the main flux rope is approximately delineated by the dashed grey and pink lines in the cases of slow and faster solar wind, respectively. During propagation to 1 AU, some interesting differences form in the plasma pressure gradient, and the background level begins to even out between the two cases due to the large distances and density decrease. The plasma accumulation in front of the CME in each simulation creates a peak that is present only in $-\nabla P_P$. We investigated whether this structure is a shock by calculating the MHD wave velocities of the background solar winds. These velocities are shown in Fig. \ref{fig:wave_speeds}. The Alfv\'en and sonic speeds are not shown because they become almost identical to the slow and fast magnetoacoustic velocities at the large distances that we are interested in. We then calculated the speed of the structure from the average of four snapshots with the peak around 1 AU, and subtracted the speed of the background wind for each case. The two asterisks in  Fig. \ref{fig:wave_speeds} indicate this difference in speed at 1 AU for the two single eruption cases.  Interestingly, they are both higher than the slow and fast magnetoacoustic velocities of their respective background winds. This implies that during the propagation of the CMEs to Earth, the leading edge of the plasma accumulation became a fast shock, not necessarily because the plasma significantly accelerated, but because of the decrease in the wave velocities. For a quantitative assessment, the speeds of the shock and the respective background wind are 387 km s$^{-1}$ and 339 km s$^{-1}$ (slow wind), and 423 km s$^{-1}$ and 384 km s$^{-1}$ (faster wind). This even agrees with observations of slow CMEs producing shocks reported by \citet{liu_2016} and \citet{lugaz_2017}, for example. The plasma pressure gradient in Fig. \ref{fig:plasma_press_grad_single_er} inside the flux rope is similar for both cases, between $\approx$ 220 R$_{\sun}$ and 195 R$_{\sun}$. However, the CME in the faster-wind case is more compressed and is followed by a depression in $-\nabla P_P$ and a reconfiguration of the equatorial current sheet below 160 R$_{\sun}$. The CME in the slow-wind case is larger and not as well structured in the tail because of the trailing plasma blobs.

\begin{figure}
	\centering
	\begin{overpic}[width=0.5\textwidth, trim={0 0 12cm 0}, clip]{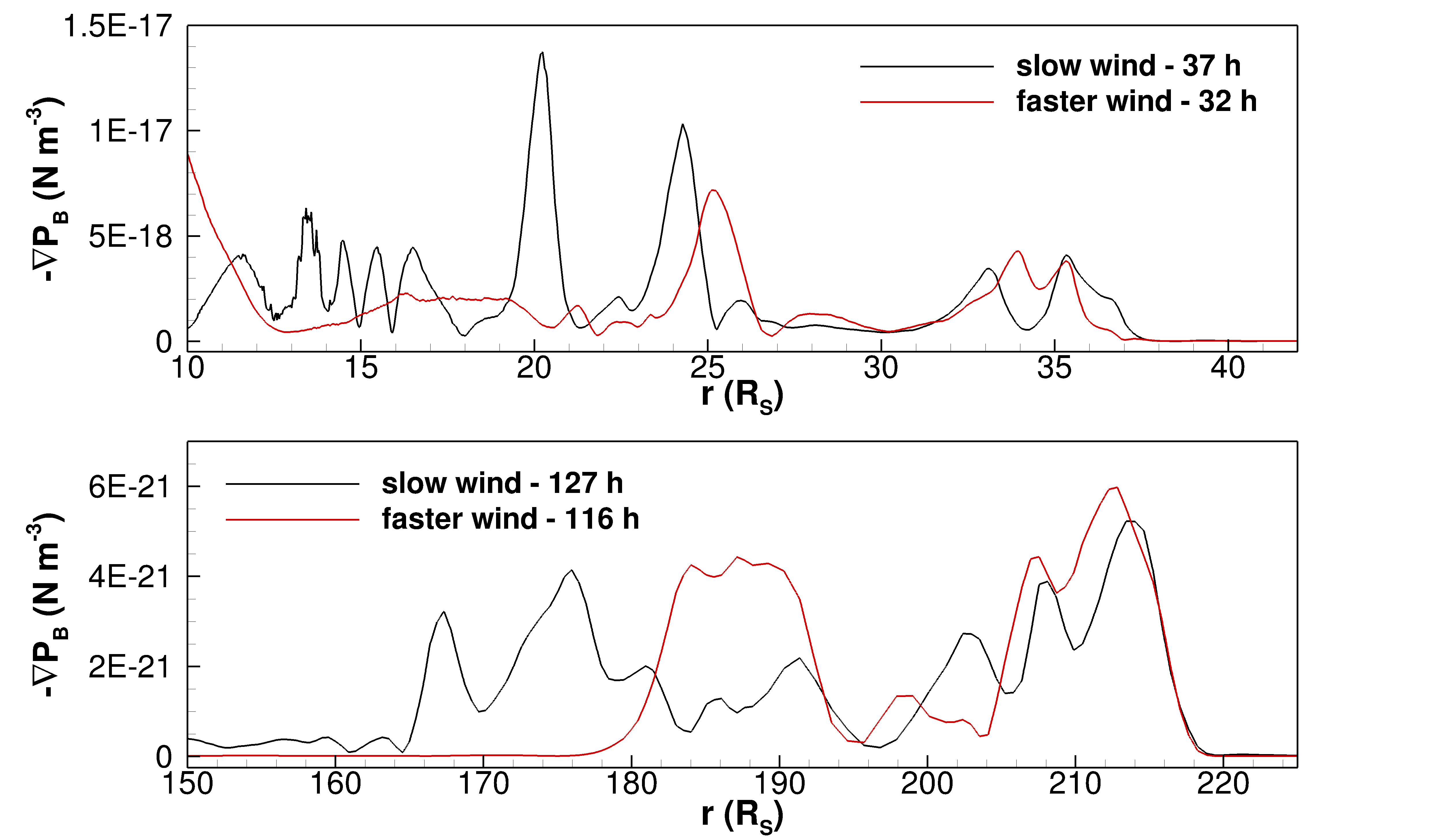}
		\put(0,0){\color{gray} \thicklines \dashline[30]{1.5}(78.6,37)(78.6,60)}
		\put(0,0){\color{gray} \thicklines \dashline[30]{1.5}(50.2,37)(50.2,60)}
		\put(0,0){\color{pink} \thicklines \dashline[30]{1.5}(78.3,37)(78.3,60)}
		\put(0,0){\color{pink} \thicklines \dashline[30]{1.5}(52,37)(52,60)}
		
		\put(0,0){\color{gray} \thicklines \dashline[30]{1.5}(89.1,6)(89.1,29)}
		\put(0,0){\color{gray} \thicklines \dashline[30]{1.5}(36,6)(36,29)}
		\put(0,0){\color{pink} \thicklines \dashline[30]{1.5}(88.8,6)(88.8,29)}
		\put(0,0){\color{pink} \thicklines \dashline[30]{1.5}(46.8,6)(46.8,29)} 
	\end{overpic}                
	\caption{Same as Fig. \ref{fig:plasma_press_grad_single_er}, but depicting magnetic pressure gradient values.}
	\label{fig:mag_press_grad_single_er}
	%       \vspace{-15px}  
\end{figure}

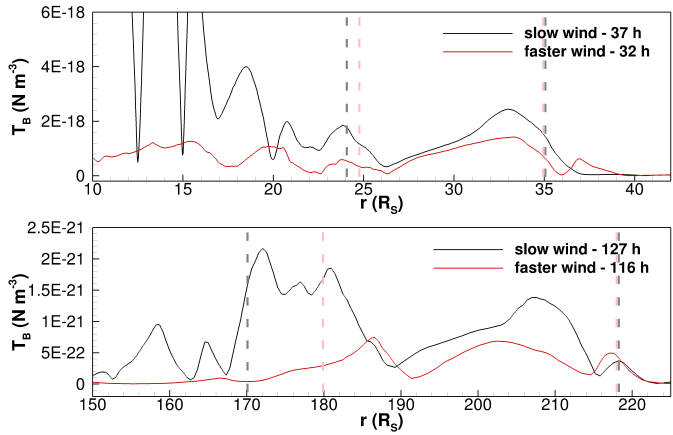
\begin{figure}
	\centering
	\begin{overpic}[width=0.5\textwidth, trim={0 0 12cm 0}, clip]{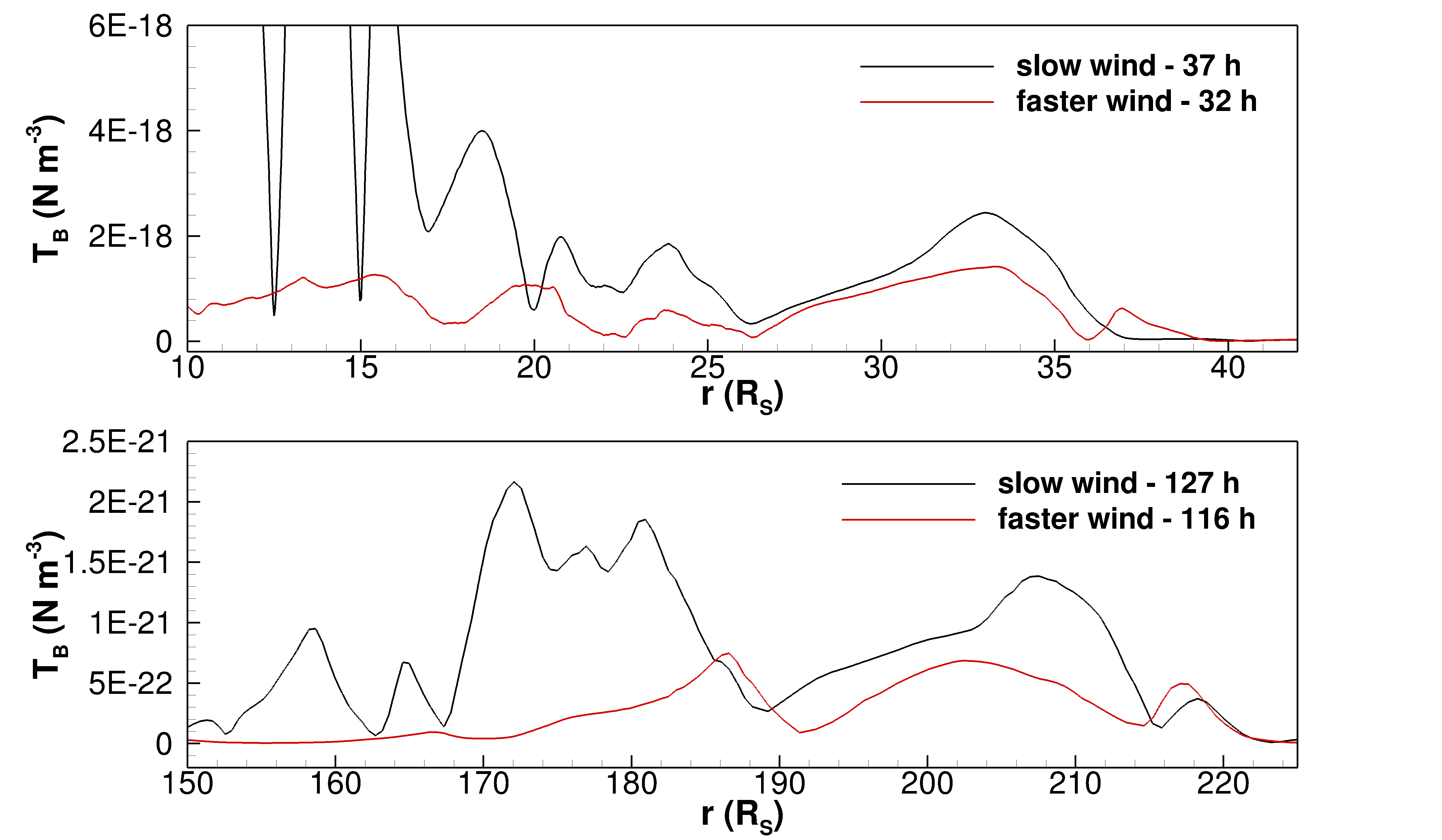}
		\put(0,0){\color{gray} \thicklines \dashline[30]{1.5}(78.6,37)(78.6,60)}
		\put(0,0){\color{gray} \thicklines \dashline[30]{1.5}(50.2,37)(50.2,60)}
		\put(0,0){\color{pink} \thicklines \dashline[30]{1.5}(78.3,37)(78.3,60)}
		\put(0,0){\color{pink} \thicklines \dashline[30]{1.5}(52,37)(52,60)}
		
		\put(0,0){\color{gray} \thicklines \dashline[30]{1.5}(89.1,6)(89.1,29)}
		\put(0,0){\color{gray} \thicklines \dashline[30]{1.5}(36,6)(36,29)}
		\put(0,0){\color{pink} \thicklines \dashline[30]{1.5}(88.8,6)(88.8,29)}
		\put(0,0){\color{pink} \thicklines \dashline[30]{1.5}(46.8,6)(46.8,29)} 
	\end{overpic}        
	\caption{Same as Fig. \ref{fig:plasma_press_grad_single_er}, but depicting magnetic tension values.}
	\label{fig:mag_tension_single_er}
	%       \vspace{-15px}  
\end{figure}

The magnetic pressure gradient shown in Fig. \ref{fig:mag_press_grad_single_er} shows two peaks at the front of the flux rope in both cases. This contrasts with the CME structure simulated and analysed by \citet{skralan_forces}, which had only one peak at about the same distance from the Sun (at $\approx$ 35 R$_{\sun}$). The first increase seen in our simulations can be attributed to the overlying magnetic field at the eruption time and to the streamer that is stripped away along with the CME. This would not be present in the case of \citet{skralan_forces}, who just injected a plasma blob into the solar wind and did not model the eruption. This suggests that different triggering mechanisms can affect the structure of CMEs, including during propagation, and therefore should be taken into consideration when interpreting CME structure. The sunward edge of the flux rope is clearly distinguishable in $-\nabla P_B$ in the faster-wind case both close to the Sun (located at 25 R$_{\sun}$) and also at 180 R$_{\sun}$ - 194 R$_{\sun}$. On the other hand, the trailing part of the CME in the slow wind is more complex: the flux rope and the trailing region ending at 24 R$_{\sun}$ and 20 R$_{\sun}$, respectively, are followed by several peaks indicating the plasma blobs. During propagation to Earth, these trailing peaks combine into only two peaks as the structure becomes more similar to that of the faster-wind case.\\
The magnetic tension shown in Fig. \ref{fig:mag_tension_single_er} makes the smallest contribution to the total force, except in the region of the trailing plasma blobs when close to the Sun. At 1 AU, the blobs merge and create a plateau in $\mathbf{T}_B$ in the aftermath of the CME. An interesting aspect is that $\mathbf{T}_B$ is consistently higher in the slow-wind case than in the faster-wind case throughout the propagation. This also slowed the eruption close to the Sun in the slow-wind case, as previously discussed in Subsection \ref{subsec:faster cme in faster wind}.

        %%-------------------------------------------------------------------   
        
        \section{Summary} \label{sec:summary}  
This paper completes the analysis, discussed in two previous papers (paper~I and paper~II), involving numerical MHD simulations of CMEs triggered by boundary shearing motions. We studied here the CMEs generated by five different erupting scenarios by calculating the forces governing their dynamics. Three of the eruptions were propagated into a slow background solar wind, and two into a faster solar wind. We addressed four aspects of the eruptions that mostly arose from the differences in background wind and consequently, in the initial magnetic configuration.\\
Firstly, we traced the cause of the faster eruption of CMEs in the faster solar wind cases (when the same shearing speed was applied as in the slow-wind cases) to the more closed streamer structure in the slow-wind cases. In our simulations, the CMEs erupt from the southernmost arcade and are deflected towards the equatorial plane. In this process, there is significantly more reconnection between the ejected flux ropes and the large northern arcade in the slow-wind scenarios. Therefore, our conclusion is that in the case of slow CMEs, the adjacent magnetic structures are important because they can decelerate the eruption via magnetic reconnection and create overlying loops that increase the confining downward magnetic tension. This interpretation also accounts for why in the stealth speed case two flux ropes erupt from the shearing velocity, whereas the same shear applied in the slow-wind case results in one CME and one non-eruptive flux rope.\\
Secondly, we explained the formation of the stealth ejecta by calculating the forces in the region of the current sheet created in the aftermath of CME1. As CME1 is ejected, the flux rope formed by the imposed shearing motions falls back to the Sun because of the insufficient energy build-up. This process stretches the current sheet and the magnetic pressure gradient pushes the sides of the magnetic field lines, pinching the current sheet. This creates a flux rope (stealth ejecta) that is ejected onto a similar path as CME1.\\
Thirdly, plasma blobs detaching from the streamer cusp in the aftermath of eruptions are also caused by the different initial magnetic structure. In the slow-wind case, the northern arcade is greatly compressed by the deflected CMEs, and in the process of returning to its original state, it elongates. The $-\nabla P_B$ forces pressing from both sides of the current sheet then induce magnetic reconnection, which creates the blobs.\\ 
Finally, we analysed the interaction of the two CMEs in the double eruption case. We were able to distinguish the main regions of the two flux ropes, and, through analysis of the forces present, we explained their merging as well as their pancaking. Furthermore, we studied the effect of the two background solar winds in the two similar single eruption cases. Surprisingly, even though the CMEs are slow, the plasma accumulated ahead of the CMEs creates a peak in $-\nabla P_P$ that evolves into a fast shock by the time they reach Earth. The double-peak structure of $-\nabla P_B$ in front of the CME could indicate that signatures of the processes triggering CMEs are retained by CMEs as they propagate through the solar wind. This suggests that CME triggering should be taken into consideration when interpreting CME structure from in-situ observations. Magnetic tension is higher in the slow solar wind case than for the faster solar wind, and it contributes to the initial deceleration of the CME. However, it makes the smallest contribution to the CME dynamics in interplanetary space.\par
This research concludes the study presented in papers~I and II, which was initiated by an observed MCME event. In papers I and II, we performed and analysed MHD numerical simulations consistent with the observations, and explored the influence of different shearing motions on the dynamics of the resulting eruptions. We then propagated these eruptions through two different background solar winds, tracked their evolution via in-situ signatures, and computed their geoeffectiveness. In the current paper, we explained the formation and dynamics of some particular features present in the simulations by computing the three main forces ($-\nabla P_P$, $-\nabla P_B$ and $\mathbf{T}_B$), and by assessing their contributions to the overall kinematics of the CMEs. This technique of force vector decomposition has to our knowledge been applied extensively to 2.5D MHD simulations for the first time here. This study demonstrates that this technique is a valuable tool for research on CME eruption and propagation through the solar wind.  

        \begin{acknowledgements}
        
        We thank the referee for the useful, constructive and comprehensive suggestions that helped to improve the manuscript. We thank Nicolas Wijsen for the insightful discussions. D.C.T. was funded by the Ph.D. fellowship of the Research Foundation – Flanders (FWO), contract number T1 1118918N. D.C.T., E.D. and M.M. acknowledge support from the Belgian Federal Science Policy Office (BELSPO) in the framework of the ESA-PRODEX program, grant No. 4000120800. I.G.R. acknowledges support from NASA program NNH17ZDA001N-LWS. This work was partially supported by a PhD Grant awarded by the Royal Observatory of Belgium. This research has received funding from the European Union's Horizon 2020 research and innovation programme under grant agreement No 870405 (EUHFORIA 2.0) and the ESA project "Heliospheric modelling techniques" (Contract No. 4000133080/20/NL/CRS). These results were also obtained in the framework of the projects C14/19/089 (C1 project Internal Funds KU Leuven), G.0D07.19N (FWO-Vlaanderen), SIDC Data Exploitation (ESA Prodex-12), and BELSPO projects BR/165/A2/CCSOM and B2/191/P1/SWiM. For the computations, we used the infrastructure of the VSC – Flemish Supercomputer Center, funded by the Hercules foundation and the Flemish Government – department EWI.

        \end{acknowledgements}

        \bibliographystyle{aa}
        \bibliography{talpeanu}
        
\end{document}